\documentclass[12pt,a4paper]{article}

\usepackage{amsmath,bm}
\usepackage{amssymb}
\usepackage{amsfonts}  
\usepackage[vcentermath]{youngtab}
\usepackage{jheppub}

\makeatletter
\g@addto@macro\bfseries{\boldmath}
\makeatother

\title{The symmetric orbifold of ${\cal N}=2$ minimal
models}
\author{Matthias~R.~Gaberdiel} 
\author{and Maximilian~Kelm}
\affiliation{Institut f\"ur Theoretische Physik, ETH Z\"urich, \\
CH-8093 Z\"urich, Switzerland}

\emailAdd{gaberdiel@itp.phys.ethz.ch} \emailAdd{mkelm@itp.phys.ethz.ch}

\abstract{The large level limit of the ${\cal N}=2$ minimal models that appear in the duality with the 
${\cal N}=2$ supersymmetric higher spin theory on AdS$_3$ is shown to be a natural subsector of a 
certain symmetric orbifold theory. We study the relevant decompositions in both the untwisted and
the twisted sector, and analyse the structure of the higher spin representations in the twisted sector
in some detail. These results should help to identify the string background of which the higher spin
theory is expected to describe the leading Regge trajectory in the tensionless limit.}

\newcommand{\abs}[1]{\lvert #1 \rvert}
\newcommand{\ysm}{\Yboxdim{5pt}}
\newcommand{\ytext}{\Yboxdim{8pt}}
\newcommand{\normord}[1]{{:}\!\mathrel{#1}\!{:}}

\DeclareMathOperator{\Tr}{Tr}

\numberwithin{equation}{section}
\allowdisplaybreaks

\def\be{\begin{equation}}
\def\ee{\end{equation}}

\def\nn{\nonumber}

\begin{document}
\maketitle

\section{Introduction}

During the last few years fairly concrete evidence has emerged for the idea that Vasiliev higher spin theories
\cite{Vasiliev:2003ev} arise as classically consistent subtheories of string theory in the tensionless limit, as had been
anticipated many years ago \cite{Sundborg:2000wp,Witten,Mikhailov:2002bp}. In particular, a
relation of this kind was suggested for the case of AdS$_4$ in \cite{Chang:2012kt}, while for 
AdS$_3$ a somewhat different proposal was made in \cite{Gaberdiel:2014cha}. In the latter case, the ${\cal N}=4$
superconformal generalisation \cite{Gaberdiel:2013vva} of the original bosonic minimal model holography of \cite{Gaberdiel:2010pz},
relating a higher spin theory on AdS$_3$ \cite{Prokushkin:1998bq,Prokushkin:1998vn} to the large $N$ limit of a 
family of minimal model CFTs, was shown to define a subtheory  of the CFT dual of string theory. 
More specifically, 
this was only shown for the background of the form AdS$_3 \times {\rm S}^3 \times \mathbb{T}^4$, where the CFT dual
of string theory is believed to be described by the symmetric orbifold of $\mathbb{T}^4$, see \cite{David:2002wn} for a review. 
The CFT duals of the ${\cal N}=4$ higher spin theories on AdS$_3$ are described by the so-called Wolf space cosets,
see \cite{Sevrin:1988ew,Schoutens:1988ig,Spindel:1988sr,Goddard:1988wv,VanProeyen:1989me,Sevrin:1989ce} for some 
early literature on this subject; in the limit where the torus background is approached --- this is the case where the level $k$
of the cosets is taken to infinity --- these cosets simplify to become the theory of $4 (N+1)$ free bosons and fermions, 
subject to a ${\rm U}(N)$ singlet constraint. They then form a natural subsector of the untwisted sector of the 
symmetric orbifold where the same free theory is only subjected to a singlet constraint under the permutation group 
$S_{N+1}\subset  {\rm U}(N)$. 
\smallskip

It is obviously tempting to believe that this sort of relation is not just restricted to the maximally supersymmetric
setting, but that the less-supersymmetric higher spin -- CFT dualities may also be related naturally to string theory. One 
particularly interesting case is the ${\cal N}=2$ version of the duality \cite{Creutzig:2011fe,Candu:2012jq}, 
for which the dual 2d CFTs are Kazama-Suzuki (KS) models \cite{Kazama:1988qp,Kazama:1988uz} that have 
an additional parameter and may therefore allow for a matrix-like construction as in \cite{Chang:2012kt}, see
\cite{Creutzig:2014ula} for an attempt in this direction. In this paper we follow a different route by trying to imitate the 
analysis of \cite{Gaberdiel:2014cha} for the ${\cal N}=2$ case: following on from our earlier work \cite{Gaberdiel:2014vca}
(see also \cite{Fredenhagen:2012bw,Fredenhagen:2014kia}),
where we showed that the large level limit of the relevant KS models can be described as the continuous orbifold 
of a free theory, we discuss how 
this (constrained) free theory is related to a symmetric orbifold construction. This symmetric orbifold is quite plausibly dual to string theory
on AdS$_3$, following the general philosophy of \cite{Hartman:2014oaa}, see also \cite{Haehl:2014yla,Belin:2014fna,Belin:2015hwa} for 
subsequent work. 
\smallskip

The paper is organised as follows. In Section~2 we define the symmetric orbifold in question, and explain how the
large level limit of the relevant KS models describe a sub-sector of this theory. In particular, we study the embedding in
detail for the untwisted sector, where we can give very concrete decompositions in terms of the representations
of the ${\cal N}=2$ $s{\cal W}_\infty$ algebra. Section~3 is devoted to understanding how the twisted sector states of the symmetric orbifold
can be similarly described in terms of these representations; we study in detail the $(2)$-cycle, as well as the 
$(2)^2$-cycle twisted sector, for which we give detailed decomposition formulae; we also explain how the
structure of a general twisted sector can be understood in similar terms. Finally, we undertake (in Section~3.4) 
first steps towards characterising the higher spin representations that are relevant for the description of the twisted
sector, generalising the recent discussion of \cite{Gaberdiel:2015wpo} to the ${\cal N}=2$ case. We end with some
conclusions, and there is an appendix where a self-contained description of the
 low-lying bosonic generators of the $s{\cal W}_\infty$ 
algebra in terms of the KS cosets is given. (This analysis is an important ingredient for the identification of the higher spin 
algebra 
representations, but it may also be useful in other contexts.)

\section{The untwisted sector of the symmetric orbifold}

It was shown in \cite{Gaberdiel:2014vca} that the ${\cal N}=2$ superconformal cosets that appear in the 
duality to the ${\cal N}=2$ supersymmetric higher spin theory on AdS$_3$ 
can be expressed as a continuous orbifold of a free field theory in the limit where the level $k\rightarrow \infty$. More precisely, 
in this limit the coset 
(see \cite{Candu:2012jq} for our conventions)
\be
\label{eq:coset}
\frac{\mathfrak{su}(N+1)^{(1)}_{k+N+1}}{\mathfrak{su}(N)^{(1)}_{k+N+1} \oplus \mathfrak{u}(1)^{(1)}_\kappa} 
\ \cong \ \frac{\mathfrak{su}(N+1)_{k}\oplus \mathfrak{so}(2N)_1}{\mathfrak{su}(N)_{k+1} \oplus \mathfrak{u}(1)_\kappa}  
\ee
was shown to agree with an orbifold theory of $N$ free bosons and fermions by the continuous
orbifold group $\mathrm{U}(N)$.  A similar 
approach was applied to the $\mathcal{N}=4$ Wolf space cosets in \cite{Gaberdiel:2014cha},
where it was shown that the corresponding coset algebra is a natural subalgebra of the chiral algebra of the 
symmetric orbifold; in turn the symmetric orbifold is believed to be  
dual to string theory on AdS$_3$, thus exhibiting how the higher spin theory is embedded into string theory. In 
this paper we want to analyse how the ${\cal N}=2$ cosets \eqref{eq:coset} can be related to an 
$\mathcal{N}=2$ symmetric orbifold. This should be a first step towards understanding the string theory 
interpretation of the corresponding ${\cal N}=2$ higher spin theory. 
\smallskip

The continuous orbifold describes the theory of $N$ free complex bosons and fermions
transforming in the fundamental (and anti-fundamental) representation of $\mathrm{U}(N)$. 
Thus it can be represented as the orbifold $(\mathbb{T}^2)^{N}/\mathrm{U}(N)$.\footnote{Strictly
speaking the relevant orbifold is $(\mathbb{R}^2)^{N} / \mathrm{U}(N)$, since the ${\rm U}(N)$ action 
is not compatible with discrete momenta. However, we shall usually refer to it as the torus orbifold since the zero 
momentum sector (which is what we shall be considering) is independent of the radius of the torus.}
The untwisted sector consists of the states that are 
invariant under the action of $\mathrm{U}(N)$. 
The full orbifold theory
includes also a twisted sector for each conjugacy class
of $\mathrm{U}(N)$. The conjugacy classes
can be labelled by ascending $N$-tuples $[\xi_1,\ldots,\xi_{N}]$ where
$-1/2 \leq \xi_1 \leq \cdots \leq \xi_{N}< 1/2$; the relevant conjugacy class contains then the diagonal matrix 
with eigenvalues $\exp(2\pi i \xi_l)$. The $\xi_l$, $l=1,\ldots,N$, can be interpreted as the twists of the 
$N$ free bosons and fermions.

As in the ${\cal N}=4$ case one may then consider, instead of the ${\rm U}(N)$ action, the permutation action of  
$S_{N+1} \subset {\rm U}(N)$. To explain this, it is natural to start with a theory of $N+1$ 
free bosons and fermions, on which $S_{N+1}$ acts by permutations. This action is not irreducible 
since the sum of all bosons (or fermions) is invariant under the permutation action,
\begin{equation}\label{standard}
N+1 \cong N \oplus 1\ .
\end{equation}
Here and in the following, normal font is used to denote representations of $S_{N+1}$,
while bold font is reserved for representations of $\mathrm{U}(N)$. The $N$-dimensional
representation on the right hand side is irreducible and is called the standard representation
of $S_{N+1}$. In a suitable basis this representation acts on only $N$ copies of $\mathbb{T}^2$, 
so the orbifold of $N+1$ copies decomposes in fact as
\begin{equation}\label{eq:torus_decomp}
 (\mathbb{T}^2)^{N+1}/S_{N+1} \cong (\mathbb{T}^2)^{N}/S_{N+1} \oplus \mathbb{T}^2\ .
\end{equation}
The free torus which transforms as a singlet under $S_{N+1}$ is not of much interest to
us and we will often drop it; in the following we shall therefore mainly concentrate on the 
non-trivial part of the symmetric orbifold. This will be the symmetric orbifold theory which
will be related to the KS models.

In order to see the relation to the KS models we recall that the standard representation
$\rho$ of $S_{N+1}$ acting on the $N$ tori maps
permutations to unitary (actually even orthogonal) $N \times N$ matrices.
Thus we can view $\rho(S_{N+1})$ as a finite
subgroup of $\mathrm{U}(N)$, and since the standard representation is faithful, 
that subgroup is isomorphic to $S_{N+1}$. Furthermore, as discussed in \cite{Gaberdiel:2014cha}, 
the fundamental (and anti-fundamental)
representation of ${\rm U}(N)$ branches down to the standard representation of $S_{N+1}$. Thus
the ${\rm U}(N)$-invariant states of the free theory form a consistent subsector of the $S_{N+1}$
invariant states, and hence the untwisted sector of the continuous orbifold is a subsector of the
untwisted sector of the symmetric orbifold. 

In the rest of this section we shall analyse the untwisted sector of the symmetric orbifold
from the viewpoint of the continuous orbifold. The twisted sectors of the symmetric orbifold
will be discussed in the following section.

\subsection{Perturbative decomposition of the untwisted sector}

The untwisted sector of the symmetric orbifold by $S_{N+1}$ contributes to the partition function as 
\begin{equation}\label{ZU}
 Z_{\text{U}}(q,\bar{q},y,\bar{y})=|\mathcal{Z}_{\text{vac}}(q,y)|^2
+\sum_R |\mathcal{Z}_R^\text{(U)}(q,y)|^2\ ,
\end{equation}
where $\mathcal{Z}_\text{vac}$ denotes the vacuum character, and $R$ labels
the non-trivial irreducible representations of $S_{N+1}$. In order to avoid having to 
write repeatedly $N+1$, we now change notation and replace the $N$ from \eqref{eq:coset}, \eqref{standard} and
\eqref{eq:torus_decomp} by $\tilde{N}$, and define $N \equiv \tilde{N}+1$; in any case, we 
shall always be considering the large $N$ (and hence large $\tilde{N}$ limit) for which this distinction 
is immaterial.
In their analysis \cite{Dijkgraaf:1996xw}, Dijkgraaf, Moore, Verlinde and Verlinde 
computed the partition function of
the symmetric orbifold $X^N/S_N$ in the R-R sector with insertion
of $(-1)^{F+\bar{F}}$, which reads
\begin{equation}
\label{eq:gen_fct_orb_RR}
 \sum_{N=0}^{\infty}p^N \tilde{Z}_\text{R}(S^N(X))=
\prod_{m=1}^{\infty}\prod_{\Delta,\bar{\Delta},\ell,\bar{\ell}}
\frac{1}{\left(1-p^m 
q^{\frac{\Delta}{m}}
\bar{q}^{\frac{\bar\Delta}{m}}
y^{\ell}\bar{y}^{\bar{\ell}}
\right)^{c(\Delta,\bar\Delta,\ell,\bar\ell)}}\ .
\end{equation}
Here we have indicated by the tilde that we have inserted a factor
of $(-1)^{F+\bar{F}}$, and 
$c(\Delta,\bar\Delta,\ell,\bar\ell)$ are the expansion coefficients of
the R-R partition function (with insertion of $(-1)^{F+\bar F}$) of the base
manifold $X$,
\begin{equation}
 \tilde{Z}_\text{R}(X)=\sum_{\Delta,\bar\Delta,\ell,\bar\ell}c(\Delta,\bar\Delta,\ell,\bar\ell)\, 
 q^{\Delta}\bar{q}^{\bar\Delta}y^{\ell}\bar{y}^{\bar\ell}\ .
\end{equation}
In our case, $X=\mathbb{T}^2$ and the partition function factorises into
its chiral parts, with 
$c(\Delta,\bar\Delta,\ell,\bar\ell)
=c(\Delta,\ell)c(\bar\Delta,\bar\ell)$.
The chiral partition function reads (as in \cite{Gaberdiel:2014cha} we will be ignoring the momentum and winding states)
\begin{align}\nonumber
 \tilde{Z}^\text{(chiral)}_\text{R}(\mathbb{T}^2)
&=i\frac{\vartheta_1(z|\tau)}{\eta^3(\tau)}
 =-(y^{\frac{1}{2}}-y^{-\frac{1}{2}})
 \prod_{n=1}^\infty\frac{(1-y q^n)(1-y^{-1} q^n)}
 {(1-q^n)^2}\\\nonumber
&= -y^{\frac{1}{2}}+y^{-\frac{1}{2}}+q(y^{\frac{3}{2}}-3y^{\frac{1}{2}}
+3y^{-\frac{1}{2}}-y^{-\frac{3}{2}})\\\nonumber
&\quad{}+3\,q^2(y^{\frac{3}{2}}-3y^{\frac{1}{2}}+3y^{-\frac{1}{2}}-y^{-\frac{3}{2}})\\
&\quad{}+q^3(-y^{\frac{5}{2}}+9y^{\frac{3}{2}}-22y^{\frac{1}{2}}+22y^{-\frac{1}{2}}-9y^{-\frac{3}{2}}+y^{-\frac{5}{2}}) + 
\mathcal{O}(q^4)\ ,
\end{align}
where
\begin{equation}
\vartheta_1(z|\tau)=i(y^{\frac{1}{2}}-y^{-\frac{1}{2}})
q^\frac{1}{8}\prod_{n=1}^\infty 
(1-q^n)(1-y\,q^n)(1-y^{-1}\,q^n)\ .
\end{equation}
In our analysis we will only be concerned with the NS-NS sector. 
The partition function in that sector can be obtained
from \eqref{eq:gen_fct_orb_RR} by spectral flow
\begin{equation}
 \label{eq:spectral_flow}
y\to y\,q^{\frac{1}{2}}\ ,\qquad \bar{y}\to \bar{y}\,\bar{q}^{\frac{1}{2}}\ ,\qquad
p\to p \,q^{\frac{1}{8}}\,\bar{q}^\frac{1}{8}\,y^\frac{1}{2}\,\bar{y}^\frac{1}{2}\ .
\end{equation}
This leads to an overall factor of 
$(q\bar{q})^{-\frac{N}{8}}=(q\bar{q})^{-\frac{c}{24}}$,
which we will suppress throughout this paper
for better readability. (Effectively, this is equivalent to multiplying the right-hand side
of the last replacement in (\ref{eq:spectral_flow})  by an additional
factor of $(q\bar{q})^{\frac{1}{8}}$.)
We then obtain the symmetric orbifold generating function in the NS-NS sector
(without a $(-1)^{F+\bar{F}}$ insertion)
\begin{equation}
\label{eq:gen_fct_orb}
 \sum_{N=0}^{\infty}p^N Z(S^N(X))=
\prod_{m=1}^{\infty}\prod_{\substack{\Delta,\bar{\Delta}\\\ell,\bar{\ell}}}
\frac{1}{\left(1-(-1)^{\ell+\bar\ell+1}p^m 
q^{\frac{\Delta}{m}+\frac{\ell}{2}+\frac{m}{4}}
\bar{q}^{\frac{\bar\Delta}{m}+\frac{\bar\ell}{2}+\frac{m}{4}}
y^{\ell+\frac{m}{2}}\bar{y}^{\bar{\ell}+\frac{m}{2}}
\right)^{c(\Delta,\bar\Delta,\ell,\bar\ell)}}\ .
\end{equation}
Now the generating function of the untwisted sector corresponds
to the $m=1$ factor of \eqref{eq:gen_fct_orb},
\begin{equation}
\label{eq:gen_fct}
 \sum_{N=0}^{\infty}p^N Z^\text{(U)}(S^N(X))=
\prod_{\substack{\Delta,\bar{\Delta}\\\ell,\bar{\ell}}}
\frac{1}{\left(1-(-1)^{\ell+\bar\ell+1}p \,
q^{\Delta+\frac{\ell}{2}+\frac{1}{4}}
\bar{q}^{\bar\Delta+\frac{\bar\ell}{2}+\frac{1}{4}}
y^{\ell+\frac{1}{2}}\bar{y}^{\bar{\ell}+\frac{1}{2}}
\right)^{c(\Delta,\bar\Delta,\ell,\bar\ell)}}\ ,
\end{equation}
and the chiral vacuum character (the partition function of the $\mathcal{W}$ algebra)
of the orbifold $(\mathbb{T}^2)^{\tilde{N}+1}/S_{\tilde{N}+1}$ can be found from \eqref{eq:gen_fct} by setting $\bar\Delta=0, \bar\ell=-\frac{1}{2}$
and taking $N$ large enough so that the coefficients stabilise; it is given by
\begin{align}
\mathcal{Z}_\text{vac}'&=1+q^{\frac{1}{2}}(y+y^{-1})+4 q
  +6q^{\frac{3}{2}} ( y+y^{-1})+4q^2(y^2+6+y^{-2})\nonumber\\
  &\quad{}+q^{\frac{5}{2}}(y^3+37 y+37y^{-1}+y^{-3})
  +7q^3 (4y^2+17+4y^{-2})+\mathcal{O}(q^{\frac{7}{2}})\ .
\end{align}
In order to obtain the vacuum character of the orbifold $(\mathbb{T}^2)^{\tilde{N}}/S_{\tilde{N}+1}
\equiv (\mathbb{T}^2)^{N-1}/S_{N}$,
we have to divide this by the chiral partition function of $\mathbb{T}^2$, 
which means we neglect the torus that transforms as a singlet under $S_{N}\equiv S_{\tilde{N}+1}$
and corresponds to the trivial factor in the permutation representation of $S_{\tilde{N}+1}$,
see eq.~(\ref{standard}).
Since this torus partition function is given by
\begin{equation}
 Z^\text{(chiral)}_\text{NS}(\mathbb{T}^2)
=\prod_{n=1}^\infty\frac{(1+yq^{n-1/2})(1+y^{-1}q^{n-1/2})}
{(1-q^n)^2}\ ,
\end{equation}
where we have once more suppressed the prefactor
 $q^{-\frac{1}{8}}$, we obtain the modified vacuum character
\begin{align}\nonumber
\mathcal{Z}_\text{vac}(q,y)
&=\frac{\mathcal{Z}_\text{vac}'}
 { Z^\text{(chiral)}_\text{NS}(\mathbb{T}^2)}\\\nonumber
&=1+q+2 q^{\frac{3}{2}} (y+y^{-1})+q^2 (y^2+8 +y^{-2})+10 q^{\frac{5}{2}}
   (y+y^{-1})\\\nonumber
&\quad{}+q^3 (5 y^2+32+5y^{-2})
+q^\frac{7}{2} (2 y^3+47 y+47y^{-1}+2y^{-3})\\
&\quad{}+q^4 (y^4+37 y^2+142+37 y^{-2}+y^{-4})+\mathcal{O}(q^{\frac{9}{2}})\ . \label{2.15}
\end{align}
This vacuum character counts the chiral states that transform
trivially under $S_{N}$, and hence includes, in particular, the
character of the ${\cal N}=2$  coset $s{\cal W}_{\infty}$  algebra (in the limit $k\rightarrow \infty$).
Thus the vacuum sector should decompose into the coset characters as
\begin{equation}\label{vacdec}
\mathcal{Z}_\text{vac}(q,y)
=\sum_{\Lambda} n(\Lambda) \, \chi_{(0;\Lambda)}(q,y)\ .
\end{equation}
Indeed, by comparing both sides of the equation order by order in $q$, we find explicitly 
\begin{align}\label{eq:vac_decomp}
\mathcal{Z}_\text{vac}(q,y)&=\chi_{(0;0)}(y,q)+
\chi_{(0;[2,0,...,0])}(q,y) + \chi_{(0;[0,0,...,0,2])}(q,y) \nonumber \\
& \quad{}+ \chi_{(0;[3,0,...,0,0])}(q,y) + \chi_{(0;[0,0,0,...,0,3])}(q,y) \nonumber \\
& \quad{}+  \chi_{(0;[2,0,...,0,1])}(q,y) + \chi_{(0;[1,0,0,...,0,2])}(q,y) \nonumber \\
& \quad{}+  2 \cdot \chi_{(0;[4,0,...,0,0])}(q,y) + 2 \cdot \chi_{(0;[0,0,0,...,0,4])}(q,y) \nonumber  \\
& \quad{}+  \chi_{(0;[0,2,0,...0,0])}(q,y) +  \chi_{(0;[0,0,...0,2,0])}(q,y) \nonumber \\
& \quad{}+\chi_{(0;[3,0,...,0,1])}(q,y) + \chi_{(0;[1,0,0,...,0,3])}(q,y) \nonumber \\
& \quad{}+2\cdot \chi_{(0;[2,0,0,...,0,2])}(q,y) \nonumber  \\
& \quad{}+\chi_{(0;[2,1,0,...,0,1])}(q,y) + \chi_{(0;[1,0,...,0,1,2])}(q,y) \nonumber \\
& \quad{}+ \chi_{(0;[0,2,0,...,0,1])}(q,y) + \chi_{(0;[1,0,...,0,2,0])}(q,y) \nonumber \\
& \quad{}+3\cdot \chi_{(0;[3,0,...,0,2])}(q,y) + 3\cdot \chi_{(0;[2,0,...,0,3])}(q,y) \nonumber \\
& \quad{}+ \chi_{(0;[1,1,0,...,0,2])}(q,y) +  \chi_{(0;[2,0,...,0,1,1])}(q,y)  \nonumber \\
& \quad{}+ \chi_{(0;[3,1,0,...,0])}(q,y) +  \chi_{(0;[0,...,0,1,3])}(q,y)  \nonumber \\
& \quad{}+2\cdot \chi_{(0;[4,0,...,0,1])}(q,y) +  2\cdot\chi_{(0;[1,0,...,0,4])}(q,y)  \nonumber \\
& \quad{}+ \chi_{(0;[2,1,0,...,0,1,0])}(q,y) +  \chi_{(0;[0,1,0,...,0,1,2])}(q,y)  \nonumber \\
& \quad{}+ \chi_{(0;[1,1,0,...,0,1,1])}(q,y) + {\cal O}(q^{\frac{9}{2}}) \ . 
\end{align}
As in \cite{Gaberdiel:2014cha}, this is precisely of the 
form (\ref{vacdec}), with $n(\Lambda)$  denoting the 
multiplicity of the $S_{N}$ singlet representation in the 
${\rm U}(N-1)$ representation $\Lambda$, where
we think of $\Lambda$ as a $S_{\tilde{N}+1}\equiv S_N$ 
representation using the embedding 
$S_{\tilde{N}+1}\subset {\rm U}(\tilde{N})$.\footnote{We 
thank Marco Baggio for helping us compute these multiplicities.}

Furthermore, as in \cite{Gaberdiel:2015mra}, we can identify the single particle generators that generate
this extended ${\cal W}$-algebra; if we had not divided out by the diagonal $\mathbb{T}^2$, the generating
function of the single particle generators would have been (see \cite{Gaberdiel:2015mra})
\begin{align}
\sum_{s,l} \tilde{d}(s,l)  q^s y^l &= (1- q) 
\left[Z^\text{(chiral)}_\text{NS}(\mathbb{T}^2)(q,y) -1\right]\nonumber\\
&= (1-q)\left[ \prod_{n=1}^{\infty} 
\frac{(1 + y q^{n-1/2}) (1 + y^{-1} q^{n-1/2})}{(1-q^n)^2} -1\right]\ , 
\end{align}
where the factor of $(1-q)$ removes the derivatives, 
and $\tilde{d}(s,l)$ are the number of single particle
generators of spin $s$ and charge $l$. Dividing out 
by the diagonal torus removes just the contribution
coming from the two free fermions and bosons; 
thus the actual generating function equals 
\begin{align}\label{eq:single_part_gen}
\sum_{s,l} d(s,l)  q^s y^l & =  (1- q) 
\Bigl[ Z^\text{(chiral)}_\text{NS}(\mathbb{T}^2)(q,y) 
- \Bigl(1 + \frac{q^\frac{1}{2} ( y + y^{-1}) }{(1-q)} 
+ 2 \frac{q^1}{(1-q)} \Bigl) \Bigr]  \nonumber \\
& = q + 2 q^\frac{3}{2} (y + y^{-1}) + q^2 (6 + y^2 + y^{-2}) + 
6 q^\frac{5}{2}  (y+y^{-1}) + \cdots \ .
\end{align}
These single particle generators generate then the 
$\mathcal{W}$ algebra in the sense that
\begin{equation}
\mathcal{Z}_\text{vac}(q,y) = \prod_{s,l} \prod_{n=0}^\infty
\frac{1}{(1-y^l q^{s+n})^{(-1)^{2s}d(s,l)}}\ .
\end{equation}
They should sit in wedge representations of the ${\cal N}=2$ $s{\cal W}_{\infty}$ 
algebra, and one finds, analogously to \cite{Gaberdiel:2015mra}, 
that we have the decomposition 
\begin{equation}
\sum_{s,l} d(s,l)  q^s y^l  =  (1-q) 
\sideset{}{'}\sum_{m,n=0}^{\infty}
\chi^{({\rm wedge})}_{(0;[m,0,0,\ldots,0,0,n])}(q,y) \ , 
\end{equation}
where the prime indicates that the terms with  
$(m,n)=(0,0), (1,0), (0,1)$ are not included in the sum. 
Note that the term with $m=n=1$ accounts precisely 
for the generators of the original $s{\cal W}_{\infty}$
algebra. We have checked these identities up to order $q^{15}$, 
and it should be straightforward to prove
them using the techniques of \cite{Gaberdiel:2015mra}.
\medskip

We can similarly extract the characters corresponding 
to the second sum in (\ref{ZU}). For 
example, the representation that contains, among others, the coset states
\begin{equation} \label{0f}
 (0;\mathrm{f})\ ,\qquad (0;\bar{\mathrm{f}})\ ,
\end{equation}
is associated to $R$ being the standard representation of $S_{N}$. 
The corresponding character 
$\mathcal{Z}_1$ is obtained from the coefficient of $\bar{q}(\bar{y}+\bar{y}^{-1})$ 
in $Z^{\text{(U)}}$, from which one has to subtract the contribution from
$|\mathcal{Z}'_\text{vac}|^2$ and then divide by the torus partition function again. 
This character turns out to be given by
\begin{align}\nonumber
\mathcal{Z}_1 &= \frac{\mathcal{Z}_\text{vac}'
(Z^\text{(chiral)}_\text{NS}(\mathbb{T}^2)-1)}
 {Z^\text{(chiral)}_\text{NS}(\mathbb{T}^2)}=
\mathcal{Z}_\text{vac}'-
\mathcal{Z}_\text{vac}\\\nonumber
&=q^\frac{1}{2} (y+y^{-1})+3 q+4 q^\frac{3}{2} (y+y^{-1})
+q^2 (3 y^2+16+3y^{-2})\\\nonumber
&\quad{}+q^\frac{5}{2} (y^3+27 y+27 y^{-1}
+y^{-3})+q^3 (23 y^2+87+23y^{-2})\\\nonumber
&\quad{}+5 q^\frac{7}{2}(2 y^3+29 y+29 y^{-1}
+2y^{-3})+q^4(3 y^4+141
   y^2+433+141 y^{-2}+3y^{-4})\\
&\quad{}+\mathcal{O}(q^\frac{9}{2})\ .
\end{align}
It can be decomposed into coset characters in the $k\to\infty$ limit according to
\begin{align}
\mathcal{Z}_1(q,y) &= \chi_{(0;[1,0,\ldots,0])}(q,y)+\chi_{(0;[0,\ldots,0,1])}(q,y)\nonumber\\
&\quad{}+\chi_{(0;[2,0,\ldots,0])}(q,y)+\chi_{(0;[0,\ldots,0,2])}(q,y)\nonumber\\
&\quad{}+\chi_{(0;[1,0,\ldots,0,1])}(q,y)\nonumber\\
&\quad{}+2\cdot\chi_{(0;[3,0,\ldots,0])}(q,y)+2\cdot\chi_{(0;[0,\ldots,0,3])}(q,y)\nonumber\\
&\quad{}+\chi_{(0;[1,1,0,\ldots,0])}(q,y)+\chi_{(0;[0,\ldots,0,1,1])}(q,y)\nonumber\\
&\quad{}+2\cdot\chi_{(0;[2,0,\ldots,0,1])}(q,y)+2\cdot\chi_{(0;[1,0,\ldots,0,2])}(q,y)\nonumber\\
&\quad{}+3\cdot\chi_{(0;[4,0,\ldots,0])}(q,y)+3\cdot\chi_{(0;[0,\ldots,0,4])}(q,y)\nonumber\\
&\quad{}+2\cdot\chi_{(0;[2,1,0,\ldots,0])}(q,y)+2\cdot\chi_{(0;[0,\ldots,0,1,2])}(q,y)\nonumber\\
&\quad{}+\chi_{(0;[0,2,0,\ldots,0])}(q,y)+\chi_{(0;[0,\ldots,0,2,0])}(q,y)\nonumber\\
&\quad{}+4\cdot\chi_{(0;[3,0,\ldots,0,1])}(q,y)+4\cdot\chi_{(0;[1,0,\ldots,0,3])}(q,y)\nonumber\\
&\quad{}+2\cdot\chi_{(0;[1,1,0,\ldots,0,1])}(q,y)+2\cdot\chi_{(0;[1,0,\ldots,0,1,1])}(q,y)\nonumber\\
&\quad{}+5\cdot\chi_{(0;[2,0,\ldots,0,2])}(q,y)\nonumber\\
&\quad{}+\chi_{(0;[2,0,\ldots,0,1,0])}(q,y)+\chi_{(0;[0,1,0\ldots,0,2])}(q,y)\nonumber\\
&\quad{}+4\cdot\chi_{(0;[3,1,0,\ldots,0])}(q,y)+4\cdot\chi_{(0;[0,\ldots,0,1,3])}(q,y)\nonumber\\
&\quad{}+\chi_{(0;[0,1,1,0,\ldots,0])}(q,y)+\chi_{(0;[0,\ldots,0,1,1,0])}(q,y)\nonumber\\
&\quad{}+7\cdot\chi_{(0;[4,0,\ldots,0,1])}(q,y)+7\cdot\chi_{(0;[1,0,\ldots,0,4])}(q,y)\nonumber\\
&\quad{}+4\cdot\chi_{(0;[2,1,0,\ldots,0,1])}(q,y)+4\cdot\chi_{(0;[1,0,\ldots,0,1,2])}(q,y)\nonumber\\
&\quad{}+3\cdot\chi_{(0;[0,2,0,\ldots,0,1])}(q,y)+3\cdot\chi_{(0;[1,0,\ldots,0,2,0])}(q,y)\nonumber\\
&\quad{}+9\cdot\chi_{(0;[3,0,\ldots,0,2])}(q,y)+9\cdot\chi_{(0;[2,0,\ldots,0,3])}(q,y)\nonumber\\
&\quad{}+2\cdot\chi_{(0;[3,0,\ldots,0,1,0])}(q,y)+2\cdot\chi_{(0;[0,1,0,\ldots,0,3])}(q,y)\nonumber\\
&\quad{}+5\cdot\chi_{(0;[1,1,0,\ldots,0,2])}(q,y)+5\cdot\chi_{(0;[2,0,\ldots,0,1,1])}(q,y)\nonumber\\
&\quad{}+\chi_{(0;[1,1,0,\ldots,0,1,0])}(q,y)+\chi_{(0;[0,1,0,\ldots,0,1,1])}(q,y)\nonumber\\
&\quad{}+\chi_{(0;[2,0,1,0\ldots,0,1])}(q,y)+\chi_{(0;[1,0,\ldots,0,1,0,2])}(q,y)\nonumber\\
&\quad{}+3\cdot\chi_{(0;[2,1,0,\ldots,0,1,0])}(q,y)+3\cdot\chi_{(0;[0,1,0,\ldots,0,1,2])}(q,y)\nonumber\\
&\quad{}+\chi_{(0;[1,0,1,0,\ldots,0,2])}(q,y)+\chi_{(0;[2,0,\ldots,0,1,0,1])}(q,y)\nonumber\\
&\quad{}+6\cdot\chi_{(0;[1,1,0,\ldots,0,1,1])}(q,y)+\mathcal{O}(q^{\frac{9}{2}})\ .
\label{eq:stand_rep}
\end{align}
This time, the coefficients of the coset characters $\chi_{(0;\Lambda)}$ correspond
precisely to the multiplicity of the $(N-1)$-dimensional standard representation of $S_{N}$
inside $\Lambda$.\footnote{Once more we thank Marco Baggio for helping
us compute these multiplicities.}
This is obviously in line with the fact that the $\tilde{N} = N-1$ boson and fermion fields 
(that give rise to the representations (\ref{0f})) 
transform precisely in this representation of the permutation group.

\subsection{The building blocks of the untwisted sector}
Having identified the lowest two representations of $S_{N}$ by explicitly evaluating
the orbifold partition function order by order in $q$, 
we will now turn to a more systematic analysis
of the untwisted sector. We will show that it organises itself in terms
of multi-particle powers of the `minimal representation' $\mathcal{Z}_1$, 
in parallel to what was observed in \cite{Gaberdiel:2015wpo}.

Let us first introduce the wedge character $\chi_1$ 
pertaining to $\mathcal{Z}_1$ by stripping off the modes outside of the wedge,
\begin{equation}
 \mathcal{Z}_1 = \mathcal{Z}_\text{vac}\cdot \chi_1 
\qquad \text{or} \qquad \chi_1 =  Z^\text{(chiral)}_\text{NS}(\mathbb{T}^2)-1\ ,
\end{equation}
where $\mathcal{Z}_\text{vac}$ is the vacuum character 
(that counts the modes outside the wedge);
explicitly, we have
\begin{align}
 \chi_1(q,y) &= \sum_{\substack{(\Delta,\ell)\\ \neq (0,-\frac{1}{2})}}
\abs{c(\Delta,\ell)} \,q^{\Delta+\frac{\ell}{2}+\frac{1}{4}}
\,y^{\ell+\frac{1}{2}} \nonumber\\
&=q^\frac{1}{2}\left(y+y^{-1}\right)+3 \,q+ 
3\,q^\frac{3}{2}\left( y+y^{-1}\right)+
q^2\left(y^2+9+y^{-2}\right)+\mathcal{O}(q^\frac{5}{2})\ . 
\end{align}
Then we claim that the full partition function of the untwisted sector for $N\to\infty$ 
can be written as
\begin{equation}
\label{eq:part_untw}
 Z^{\text{(U)}}(q,\bar q,y,\bar y)= |\mathcal{Z}_\text{vac}(q,y)|^2
\left(1+\sum_{\Lambda} |\chi_{\Lambda}(q,y)|^2\right)\ ,
\end{equation}
where $\Lambda$ runs over all Young diagrams, and 
$\chi_{\Lambda}(q,y)$ is the $\Lambda$-symmetrised power of 
$\chi_1(q,y)$ given by
(see e.g.\ \cite{Lux:2010})
\begin{equation}
\label{eq:char_symm}
 \chi_{\Lambda}(q,y)=\frac{1}{m!}\sum_{\rho \in S_m}
\chi_m^{\Lambda}(\rho)\prod_{k=1}^m 
\mathcal{F}^{k-1}\chi_1\left(q^k,y^k\right)^{a_k(\rho)}\ .
\end{equation}
Here $m=|\Lambda|$ is the number of boxes of 
$\Lambda$, $\chi_m^{\Lambda}(\rho)$ is
the character of $\Lambda$ seen as an $S_m$-representation,
$a_k(\rho)$ is the number of $k$-cycles in the permutation $\rho$,
and $\mathcal{F}$ is the involutive mapping that acts on a character
or partition function by insertion of $(-1)^{F+\bar{F}}$. 
So denoting $\mathcal{F}\chi_1$ by $\tilde{\chi}_1$, the first few characters read
\begin{align}
 \chi_{\ysm\yng(1)}(q,y)&=\chi_1(q,y)\ ,\nonumber\\
\chi_{\ysm\yng(2)}(q,y)&=\frac{1}{2}\left(\chi_1(q,y)^2
+\tilde{\chi}_1(q^2,y^2)\right)\ ,\nonumber\\
\chi_{\ysm\yng(1,1)}(q,y)&=\frac{1}{2}\left(\chi_1(q,y)^2
-\tilde{\chi}_1(q^2,y^2)\right)\ ,\nonumber\\
\chi_{\ysm\yng(3)}(q,y)&=\frac{1}{6}\left(\chi_1(q,y)^3
+3\chi_1(q,y)\tilde{\chi}_1(q^2,y^2)
+2\chi_1(q^3,y^3)\right)\ ,\nonumber\\
\chi_{\ysm\yng(1,1,1)}(q,y)&=\frac{1}{6}\left(\chi_1(q,y)^3
-3\chi_1(q,y)\tilde{\chi}_1(q^2,y^2)
+2\chi_1(q^3,y^3)\right)\ ,\nonumber\\
\chi_{\ysm\yng(2,1)}(q,y)&=\frac{1}{3}\left(\chi_1(q,y)^3
-\chi_1(q^3,y^3)\right)\ . \label{2.25}
\end{align}
A proof of \eqref{eq:part_untw} will be given at the end of section~\ref{sec:gen_twist}.
We have checked agreement of eqs.~\eqref{eq:gen_fct} and \eqref{eq:part_untw}
for up to three boxes and up to order $\mathcal{O}(q^2)\mathcal{O}(\bar{q}^2)$,
which is the lowest order to which the Young diagrams $\Lambda$ with four boxes
contribute.

\section{The twisted sector}\label{sec:twisted}

The twisted sectors are labelled by conjugacy classes $[g]$ of $S_N$, 
and consist of those states which are invariant under $C^g$,
the centraliser of $g$ in $S_{N}$. 
The conjugacy classes of $S_N$ can be
labelled by cycle structures
\begin{equation}
(1)^{N_1}(2)^{N_2}(3)^{N_3}\cdots (m)^{N_m}\ ,
\qquad
\text{where }
\sum_{i=1}^m N_i = N\ .
\end{equation}
The conjugacy class labelled by such a string consists
of all elements of $S_{N}$ that can be decomposed
into $N_2$ 2-cycles, $N_3$ 3-cycles, etc. 
The centraliser of this conjugacy class is then
\begin{equation}
C^{(1)^{N_1}(2)^{N_2}\cdots (m)^{N_m}}
\cong S_{N_1}\times (S_{N_2} \ltimes \mathbb{Z}_2^{N_2})
\times \cdots \times (S_{N_m} \ltimes \mathbb{Z}_m^{N_m})\ .
\end{equation}
The $n$ free fermions and bosons corresponding to an $n$-cycle
have twists of $i/n$, for $i=1,\ldots,n$, and the corresponding
$\mathbb{Z}_n$ acts by the usual phases on them. On the other hand, 
the $S_{N_n}$ factors in the semi-direct products permute the $N_n$ 
different $n$-cycles among each other.

Since states are tensor products of
left- and right-moving states, the action of the centraliser on these chiral states need not be trivial (only
the combined action on left- and right-movers must be). 
The partition function of the $[g]$-twisted sector will thus have the structure
\begin{equation}
 Z^{[g]}= \sum_{R}|\mathcal{Z}^{[g]}_{R}|^2\ ,
\end{equation}
where $R$ labels the different irreducible representations
of the centraliser $C^{[g]}$. We will see examples of this below.

\subsection{The 2-cycle twisted sector}\label{sec:2cycle}

We will start our analysis of the twisted sector with the subsector 
corresponding to a 2-cycle twist, which is the simplest example.
The partition function of the 2-cycle twisted sector in the ordinary symmetric orbifold
can be obtained from the generating function; more specifically, the R-R sector expression can be extracted from 
the $m=1$ and $m=2$ factors of \eqref{eq:gen_fct_orb_RR},
\begin{equation}
 \sum_{N=0}^{\infty}p^N \tilde{Z}^{(2)}_\text{R}(S^N \mathbb{T}^2)
=p^2 \sideset{}{'}\sum_{\Delta,\bar\Delta,\ell,\bar\ell}
c(\Delta,\bar\Delta,\ell,\bar\ell)
 q^\frac{\Delta}{2}\bar{q}^\frac{\bar\Delta}{2}
y^{\ell}\bar{y}^{\bar\ell}\cdot
\prod_{\Delta,\bar\Delta,\ell,\bar\ell}
 \frac{1}{(1-p\,q^{\Delta}\bar{q}^{\bar\Delta}
 y^{\ell}\bar{y}^{\bar\ell})^{c(\Delta,\bar\Delta,\ell,\bar\ell)}}\ ,
\end{equation}
where the prime at the sum indicates that $\Delta-\bar\Delta$ has to be even.
Flowing to the NS-NS sector and considering the stabilising limit of large $N$ we find 
for the partition function without $(-1)^{F+\bar{F}}$ insertion
\begin{align}
\label{eq:part_fct_2twist}
Z^{(2)}(S^N \mathbb{T}^2)
&=\sideset{}{'}\sum_{\Delta,\bar\Delta,\ell,\bar\ell}
|c(\Delta,\bar\Delta,\ell,\bar\ell)|\,
 q^{\frac{1}{2}(\Delta+\ell+1)}\bar{q}^{\frac{1}{2}(\bar\Delta+\bar\ell+1)}
y^{\ell+1}\bar{y}^{\bar\ell+1}\nonumber\\
&\quad {}\times \!\!\!\!\!\!\prod_{\substack{(\Delta,\bar\Delta,\ell,\bar\ell)\\
\neq (0,0,-1/2,-1/2)}}
\frac{1}{\left(1-(-1)^{\ell+\bar\ell+1} q^{\Delta+\frac{\ell}{2}+\frac{1}{4}}
\bar{q}^{\bar\Delta+\frac{\bar\ell}{2}+\frac{1}{4}}
 y^{\ell+\frac{1}{2}}\bar{y}^{\bar\ell+\frac{1}{2}}\right)^{c(\Delta,\bar\Delta,\ell,\bar\ell)}}\ .
\end{align}
We then obtain the partition function we are interested in by dividing
by the left- and right-moving torus partition function
$Z_\text{NS}(\mathbb{T}^2)=\abs{Z^\text{(chiral)}_\text{NS}(\mathbb{T}^2)}^2$,
\begin{align}
 Z^{(2)}(q,\bar q,y,\bar y)&=
 \frac{q^\frac{1}{4}\bar{q}^\frac{1}{4}}{y^\frac{1}{2}\bar{y}^\frac{1}{2}}
\bigl[1+y\bar{y}
+(y\bar{y}^2+3y+3\bar{y}+\bar{y}^{-1})\bar{q}^\frac{1}{2} \nonumber\\
&{}\quad +(y^2\bar{y}+y+\bar{y}+y^{-1})q^\frac{1}{2}\nonumber\\
&{}\quad +  2(y^2\bar{y}^2+2y^2+5y\bar{y}+2\bar{y}^2+5+2y\bar{y}^{-1}
+2y^{-1}\bar{y}+y^{-1}\bar{y}^{-1})q^\frac{1}{2}\bar{q}^\frac{1}{2}\nonumber \\
&{}\quad +  \cdots \bigr] \ . 
\end{align}
Since the centraliser of this sector (ignoring the $N-2$ sectors 
that are not affected by the twist --- invariance
with respect to this subgroup will just guarantee 
that the remaining factors give rise to a factor equal to the 
untwisted sector $Z^\text{(U)}$ for large $N$) is simply 
$S_2\cong \mathbb{Z}_2$, there are two representations that contribute, namely
\begin{align}
 \mathcal{Z}_+(q,y) &= \mathcal{Z}_{\text{vac}}\cdot
\sum_{\Delta \text{ even},\ell}
|c(\Delta,\ell)|\,q^{\frac{1}{2}(\Delta+\ell+1)}y^{\ell+1} \nonumber\\
&=y^\frac{1}{2}q^\frac{1}{4}+(y^\frac{3}{2}+3y^{-\frac{1}{2}})q^\frac{3}{4}
+ (10y^\frac{1}{2}+3 y^{-\frac{3}{2}})q^\frac{5}{4} \nonumber\\
&\quad{}+ (12 y^\frac{3}{2}+27 y^{-\frac{1}{2}}
+y^{-\frac{5}{2}})q^\frac{7}{4}+\mathcal{O}(q^\frac{9}{4})\ ,
\end{align}
and 
\begin{align}
 \mathcal{Z}_-(q,y) &=\mathcal{Z}_{\text{vac}}\cdot
\sum_{\Delta \text{ odd},\ell}
|c(\Delta,\ell)|\,q^{\frac{1}{2}(\Delta+\ell+1)}y^{\ell+1} \nonumber\\
&=y^{-\frac{1}{2}}q^\frac{1}{4}+(3y^{\frac{1}{2}}+y^{-\frac{3}{2}})q^\frac{3}{4}
+(3 y^\frac{3}{2}+10y^{-\frac{1}{2}})q^\frac{5}{4}\nonumber\\
&\quad {}+(y^\frac{5}{2}+27 y^\frac{1}{2}
+12 y^{-\frac{3}{2}})q^\frac{7}{4}+\mathcal{O}(q^\frac{9}{4})\ .
\end{align}
Defining the wedge characters $\chi^{(2)}_{\pm}$ by
\begin{equation}\label{2cywedge}
 \mathcal{Z}_{\pm} = \mathcal{Z}_\text{vac}\cdot \chi^{(2)}_{\pm} \ ,
\end{equation}
the whole sector can then simply be written as
\begin{align}
  Z^{(2)}
&= Z^\text{(U)}\cdot\left(|\chi^{(2)}_+|^2
+|\chi^{(2)}_-|^2\right)\nonumber\\
&=|\mathcal{Z}_\text{vac}|^2
\cdot  \left(1+\sum_{\Lambda} |\chi_{\Lambda}(q,y)|^2\right)
\cdot\left(|\chi^{(2)}_+|^2
+|\chi^{(2)}_-|^2\right)\ . \label{2cycle}
\end{align}
The two wedge characters $\chi_\pm$ have the same leading $q$ behaviour, and their
lowest terms are described by the coset representations \cite{Gaberdiel:2014vca}
\begin{equation}\label{2leading}
 ([k/2,0,\ldots,0];[k/2,0,\ldots,0])\ \qquad \hbox{and} \qquad 
([k/2,0,\ldots,0];[k/2+1,0,\ldots,0])
\end{equation}
for large $k$, respectively, i.e., have twist 
$ \xi=[-1/2,0,\ldots,0]$
in the continuous orbifold picture.
One of these states can be obtained from the other
by acting on it with a fermionic zero-mode. In fact, both $\chi_\pm$ can be written
in terms of coset representations (for $k\rightarrow\infty$), and we have checked that
up to order $q^2$ we have 
\begin{align}
 \mathcal{Z}_+(q,y) &= 
\chi_{([k/2,0,\ldots,0]; [k/2 + 1,0,\ldots,0])}(q,y) + 
 \chi_{([k/2,0,\ldots,0]; [k/2 - 1,0,\ldots,0])}(q,y) \nonumber\\
 &\quad {} + \chi_{([k/2,0,\ldots,0]; [k/2 + 3,0,\ldots,0])}(q,y)  + 
 \chi_{([k/2,0,\ldots,0]; [k/2,1,0,\ldots,0])}(q,y) \nonumber\\
 &\quad {} +\chi_{([k/2,0,\ldots,0]; [k/2 + 1,0,\ldots,0,1])}(q,y)+ 
 \chi_{([k/2,0,\ldots,0]; [k/2,1,0,\ldots,0,1])}(q,y)\nonumber\\ 
 &\quad {} +\chi_{([k/2,0,\ldots,0]; [k/2-2,1,0,\ldots,0])}(q,y) + 
 \chi_{([k/2,0,\ldots,0]; [k/2 + 3,0,\ldots,0,1])}(q,y)\nonumber\\ 
 &\quad {} +\chi_{([k/2,0,\ldots,0]; [k/2 - 1,0,\ldots,0,1])}(q,y) + 
 \chi_{([k/2,0,\ldots,0]; [k/2 + 2,1,0,\ldots,0])}(q,y) \nonumber\\
 &\quad {} +2\cdot \chi_{([k/2,0,\ldots,0];[k/2 - 1,2,0,\ldots,0])}(q,y) +
 2 \cdot\chi_{([k/2,0,\ldots,0]; [k/2 + 1,0,\ldots,0,2])}(q,y) \nonumber\\
 &\quad {} +2 \cdot\chi_{([k/2,0,\ldots,0]; [k/2 -2, 1,0,\ldots,0,1])}(q,y) 
+\mathcal{O}(q^\frac{9}{4})\ ,\nonumber\\
 \mathcal{Z}_-(q,y) &=
\chi_{([k/2,0,\ldots,0]; [k/2,0,\ldots,0])}(q,y) + 
 \chi_{([k/2,0,\ldots,0]; [k/2+2,0,\ldots,0])}(q,y) \nonumber\\ 
 &\quad {} +\chi_{([k/2,0,\ldots,0];  [k/2,0,\ldots,0,1])}(q,y) + 
 \chi_{([k/2,0,\ldots,0]; [k/2-1,1,0,\ldots,0])}(q,y) \nonumber\\
&\quad {} + \chi_{([k/2,0,\ldots,0];  [k/2-1,1,0,\ldots,0,1])}(q,y) + 
 \chi_{([k/2,0,\ldots,0]; [k/2+1,1,0,\ldots,0])}(q,y) \nonumber\\
 &\quad {} +\chi_{([k/2,0,\ldots,0];  [k/2+2,0,\ldots,0,1])}(q,y) + 
 \chi_{([k/2,0,\ldots,0];  [k/2+1,1,0,\ldots,0,1])}(q,y) \nonumber\\
 &\quad {} +\chi_{([k/2,0,\ldots,0];  [k/2-2,0,\ldots,0])}(q,y) + 
 \chi_{([k/2,0,\ldots,0]; [k/2+4,0,\ldots,0])}(q,y) \nonumber\\ 
 &\quad {} +2 \cdot\chi_{([k/2,0,\ldots,0];  [k/2,0,\ldots,0,2])}(q,y) + 
 2 \cdot\chi_{([k/2,0,\ldots,0];  [k/2-2,2,0,\ldots,0])}(q,y) \nonumber\\
 &\quad {} +\chi_{([k/2,0,\ldots,0]; [k/2-3,1,0,\ldots,0])}(q,y) + 
 \chi_{([k/2,0,\ldots,0];  [k/2-2,0,\ldots,0,1])}(q,y) 
+\mathcal{O}(q^\frac{9}{4})\ .
\end{align}

As in \cite{Gaberdiel:2014cha}, we can understand
the multiplicities in these decompositions
systematically: $\mathcal{Z}_{\pm}$ contains all those coset representations
\begin{equation}
([k/2,0,\ldots,0];[k/2+l_0,\Lambda'])
\end{equation}
for which $l_0 + \sum_i \Lambda'_i$ is odd or even,
respectively.\footnote{Here we sum only over the 
first few Dynkin labels of $\Lambda'$,
such that anti-boxes and their tensor powers do not contribute
to the $\mathbb{Z}_2$ parity. Actually, we should treat $\Lambda'$
as a $\mathrm{U}(N-2)$ rather than $\mathrm{SU}(N-2)$ representation,
since an anti-box of $\mathrm{U}(N-2)$ differs from
$[0,\ldots,0,1]$ of $\mathrm{SU}(N-2)$ by its $\mathrm{U}(1)$ charge,
which we have suppressed in our notation.\label{footnote:antibox}}
This is due to the fact that 
$l_0 + \sum_i \Lambda'_i$
counts the number of twisted modes by which
the ground state 
\begin{equation}\label{3.14}
(\Lambda_+;\Lambda_-)=([k/2,0,\ldots,0];[k/2,0,\ldots,0])
\end{equation}
has been excited. Each of these twisted modes
has odd parity under the $\mathbb{Z}_2$ in the centraliser.
In addition, each state has to be invariant under the
$S_{N-2}$ factor of the centraliser --- the states 
that are not invariant are accounted
for by the middle factor in (\ref{2cycle}).  
For the boxes in the first row of $\Lambda_-$, this
is automatically true, so the overall
multiplicity with which $(\Lambda_+;\Lambda_-)$ contributes to 
$\mathcal{Z}_{\pm}$ is determined by the
multiplicity of the trivial $S_{N-2}$ representation 
inside the $\mathrm{SU}(N-2)$ representation $\Lambda'$. 
Using the (by now) standard embedding 
$S_{N-2} \subset \mathrm{U}(N-3) \subset \mathrm{SU}(N-2)$, we obtain
the decompositions
\begin{equation}
\mathbf{(N-2)}_{\mathrm{SU}(N-2)}
\to (N-3)_{S_{N-2}} \oplus 1_{S_{N-2}}\ , \quad
\mathbf{\overline{(N-2)}}_{\mathrm{SU}(N-2)}
\to (N-3)_{S_{N-2}} \oplus 1_{S_{N-2}}\ .
\end{equation}
Hence states with $\Lambda' = \ytext\yng(1)$ or 
$\Lambda' = \ytext\overline{\yng(1)}$ have multiplicity 1.
Moreover, the symmetric product of two boxes contains two
$S_{N-2}$ singlets, whereas the antisymmetric product
contains none. This explains why states with $\Lambda'=\ytext\yng(1,1)$
do not appear in the decomposition, whereas states with
$\Lambda'=\ytext\yng(2)$ appear with multiplicity $2$.
The tensor product of a box with an antibox, $\ytext\overline{\yng(1)}
\otimes \yng(1)$, contains two singlets, but one of them corresponds to the  
$s{\cal W}_\infty$ generators and hence does not give rise to a new representation;
the resulting multiplicity in the coset decomposition is therefore again $1$.

\subsection{The twisted sector with two 2-cycles}
The next, slightly more complicated step is to study 
the sector whose twist corresponds to the conjugacy class
of permutations which have two 2-cycles. 
This means that two of the free bosons and fermions are twisted,
while all the others are untwisted. 
We are interested in this sector because
it contains the operators corresponding to exactly marginal deformations
of the theory, which should, in particular, allow us to 
study the behaviour upon switching on 
the string coupling constant, compare \cite{Gaberdiel:2015uca}. By the same reasoning 
as before, we can obtain the generating function of 
the partition function from \eqref{eq:gen_fct_orb}
\begin{align}
\label{eq:twist_part_NS}
 \sum_{N=0}^{\infty}p^N Z^{(2)^2}(S^N \mathbb{T}^2)&=
\frac{p^4}{2} \Biggl[\Bigl( \sideset{}{'}\sum_{\Delta,\bar\Delta,\ell,\bar\ell}
\abs{c(\Delta,\bar\Delta,\ell,\bar\ell)}
 q^{\frac{1}{2}(\Delta+\ell+1)}\bar{q}^{\frac{1}{2}(\bar\Delta+\bar\ell+1)}
y^{\ell+1}\bar{y}^{\bar\ell+1}\Bigr)^2\nonumber\\
&\qquad\qquad{}+\sideset{}{'}\sum_{\Delta,\bar\Delta,\ell,\bar\ell}
c(\Delta,\bar\Delta,\ell,\bar\ell)
q^{\Delta+\ell+1}\bar{q}^{\bar{\Delta}
+\bar\ell+1}y^{2\ell+2}\bar{y}^{2\bar\ell+2} \Biggr]
\nonumber\\
&\quad{}\times\prod_{\Delta,\bar\Delta,\ell,\bar\ell}
 \frac{1}{(1-pq^{\Delta+\frac{\ell}{2}
 +\frac{1}{4}}\bar{q}^{\bar\Delta+\frac{\bar\ell}{2}+\frac{1}{4}}
 (-y)^{\ell+\frac{1}{2}}(-\bar{y})^{\bar\ell
 +\frac{1}{2}})^{c(\Delta,\bar\Delta,\ell,\bar\ell)}}\ .
\end{align}
In the first term, a factor of $(-1)^{\ell+\bar\ell+1}$ has again been absorbed
into $\abs{c(\Delta,\bar\Delta,\ell,\bar\ell)}$, whereas the second term contains
a factor of $(-1)^{2(\ell+\bar\ell+1)}=1$. 
As before, the partition function for our symmetric orbifold can be obtained
by taking $N$ large, and dividing by the partition function of the 
free $\mathbb{T}^2$ theory. We thus obtain
\begin{align}
 Z^{(2)^2}&=q^\frac{1}{2}\bar{q}^\frac{1}{2}(1+y\bar{y}+y^{-1}\bar{y}^{-1})
 +q^\frac{1}{2}\bar{q}^1\left(y\bar{y}+y^{-1}\bar{y}^{-1}+3(y+y^{-1})+4(\bar{y}+\bar{y}^{-1})\right)\nonumber\\
&\quad {} +q^1\bar{q}^\frac{1}{2}\left(y\bar{y}+y^{-1}\bar{y}^{-1}+4(y+y^{-1})+3(\bar{y}+\bar{y}^{-1})\right)\nonumber\\
&\quad {}+q^1\bar{q}^1 \left(38+3(y^2\bar{y}^2+y^{-2}\bar{y}^{-2})
+17(y+y^{-1})(\bar{y}+\bar{y}^{-1})\right.\nonumber\\
&\qquad\qquad \left. {}+7(y^2+\bar{y}^2+\bar{y}^{-2}+y^{-2})\right)+
\cdots\ .
\end{align}
The centraliser of this sector is
\begin{equation}
 C^{(2)^2}=S_{N-4}\times(S_2 \ltimes \mathbb{Z}_2^2)\ .
\end{equation}
Again, we can ignore the action of the $S_{N-4}$ part --- 
this will only ensure that the $N-4$ untwisted
bosons and fermions from the directions that are unaffected by the twist 
reproduce again the contribution from the untwisted sector. The remaining group 
$S_2 \ltimes \mathbb{Z}_2^2 \cong D_8$ 
(the dihedral group of order $8$) has five irreducible representations,
four of dimension $1$, and one of dimension $2$.  
In order to describe them, we first note that the 
abelian $\mathbb{Z}_2 \times \mathbb{Z}_2$ subgroup 
has $4$ different one-dimensional representations
that are characterised by the eigenvalues $(\pm,\pm)$ 
of the two non-trivial $\mathbb{Z}_2$ generators. 
In $D_8$, both $(+,+)$ and $(-,-)$ give rise to two 
one-dimensional representations each that differ by 
the sign under the exchange of $S_2$ --- this accounts 
for the $4$ one-dimensional representations.
The two-dimensional representation of $D_8$ is 
spanned by the two states with mixed charges 
$(\pm,\mp)$ that 
are exchanged under the action of $S_2$. 

The simplest way to describe the contribution of these 
representations to the twisted sector 
is in multi-particle form.
It follows from the derivation from eq.~(\ref{eq:twist_part_NS}) 
that the $(2)^2$ sector has the partition function
\begin{align}
& Z^{\text{(U)}} \cdot \frac{1}{2}\, 
\Biggl[ \biggl( \Bigl| \sum_{\substack{\Delta \text{ even,}\\ \ell}}
\abs{c(\Delta,\ell)}\,q^{\frac{1}{2}(\Delta+\ell+1)}y^{\ell+1}\Bigr|^2
+ \Bigl| \sum_{\substack{\Delta \text{ odd,}\\ \ell}}
\abs{c(\Delta,\ell)}\,q^{\frac{1}{2}(\Delta+\ell+1)}
y^{\ell+1}\Bigr|^2\biggr)^2\nonumber\\
&\qquad  \qquad {}+ \Bigl| \sum_{\substack{\Delta \text{ even,}\\ \ell}}
c(\Delta,\ell)q^{\Delta+\ell+1}y^{2\ell+2}\Bigr|^2
+ \Bigl| \sum_{\substack{\Delta \text{ odd,}\\ \ell}}
c(\Delta,\ell)q^{\Delta+\ell+1}y^{2\ell+2}\Bigr|^2\Biggr]\ .
\end{align}
Since the wedge characters of the $2$-cycle twisted sector, see eq.~(\ref{2cywedge}), are given by
\begin{equation}\label{chipm}
 \chi^{(2)}_{\pm} = \sum_{\substack{\Delta \text{ even/odd,}\\ \ell}}
\!\!\abs{c(\Delta,\ell)}\,q^{\frac{1}{2}(\Delta+\ell+1)}y^{\ell+1}\ ,
\end{equation}
the above $(2)^2$ sector partition function can then be written as 
\begin{align}
\label{eq:structure22}
 Z^{(2)^2} &=Z^{\text{(U)}} \cdot
\Bigl[\bigl|(\chi^{(2)}_{+})_{{\ysm\yng(2)}}\bigr|^2
+\bigl|(\chi^{(2)}_{+})_{{\ysm\yng(1,1)}}\bigr|^2
+\bigl|(\chi^{(2)}_{-})_{{\ysm\yng(2)}}\bigr|^2
+\bigl|(\chi^{(2)}_{-})_{{\ysm\yng(1,1)}}\bigr|^2
+\bigl|\chi^{(2)}_{+}\chi^{(2)}_{-}\bigr|^2 \Bigl]\ ,
\end{align}
where
\begin{align}
 (\chi^{(2)}_{\pm})_{{\ysm\yng(2)}/{\ysm\yng(1,1)}}(q,y)
&=\frac{1}{2}\bigl(\chi^{(2)}_{\pm}(q,y)^2
\pm \tilde{\chi}_{\pm}(q^2,y^2)\bigr)\nonumber\\
&=\frac{1}{2}\Bigl[\bigl(\!\!\!\sum_{\substack{\Delta \text{ even/odd,}\\ \ell}}
\!\!\!\abs{c(\Delta,\ell)}\,q^{\frac{1}{2}(\Delta+\ell+1)}y^{\ell+1}\bigr)^2
\pm \!\!\!\sum_{\substack{\Delta \text{ even/odd,}\\ \ell}}
\!\!\!c(\Delta,\ell)\,q^{\Delta+\ell+1}y^{2\ell+2}\Bigr]\ .
\end{align}
Each of the terms in (\ref{eq:structure22}) corresponds to one 
of the five irreducible representations
of $D_8$, and can be organised in terms of coset representations.
In order to describe this in detail, let us start from the ground state 
that has the eigenvalues $(+,+)$
with respect to the two $\mathbb{Z}_2$ factors; 
it appears in the $(\chi^{(2)}_{-})_{{\ysm\yng(1,1)}}$
sector,\footnote{Our convention for the definition of 
$\chi_\pm^{(2)}$ follows \cite{Gaberdiel:2014cha},
and is motivated by the fact that $\pm$ corresponds 
to even/odd in eq.~(\ref{chipm}); this then leads to the
somewhat strange (but inevitable) conclusion that the 
corresponding $\mathbb{Z}_2$ eigenvalue is $\mp$, 
see also eq.~(7.17) and (7.18) of \cite{Gaberdiel:2014cha}.} 
is an element of the coset representation
\begin{equation}\label{eq:22_gs}
 \bigl(\Lambda_+;\Lambda_-\bigr)=\bigl([0,k/2,0,\ldots,0];[0,k/2,0,\ldots,0]\bigr)\ ,
\end{equation}
and therefore has the twist 
$ \xi=[-1/2,-1/2,0,\ldots,0]$.
All other states of the $(2)^2$ twisted sector can be obtained by adding boxes to
$\Lambda_-$ (while leaving $\Lambda_+$ invariant), yielding
\begin{equation}
\label{eq:excited}
 \Lambda_-=[l_1,k/2+l_2,\Lambda']\ ,
\end{equation}
where $l_1,l_2 \in \mathbb{Z}$, and $\Lambda'$ denotes the remaining
$N-4$ Dynkin labels. 
For example, $l_1=0$, $l_2=1$ contains the ground state transforming as $(-,-)$ with
respect to the two $\mathbb{Z}_2$ factors --- it appears 
in the sector $(\chi^{(2)}_{+})_{{\ysm\yng(2)}}$ --- 
while $l_1=1$, $l_2=0$ contains the ground state 
with eigenvalues $(+,-)$, which appears in the
sector $\chi^{(2)}_{+}\chi^{(2)}_{-}$.
The other two dihedral representations only contribute at order $q^1$; in terms of 
coset representations we have the decompositions 
\begin{align}
\label{eq:identification22}
\mathcal{Z}_\text{vac}\cdot
(\chi^{(2)}_{+})_{{\ysm\yng(2)}} &=\chi_{([0, k/2,0,\ldots,0];[0, k/2 + 1,0,\ldots,0])} + 
    \chi_{([0, k/2,0,\ldots,0]; [2, k/2 - 1,0,\ldots,0])}\nonumber\\
  &\quad{}  + \chi_{([0, k/2,0,\ldots,0]; [0, k/2 , 1,0,\ldots,0])}  
    + \chi_{([0, k/2,0,\ldots,0]; [0, k/2 + 1,0,\ldots,0,1])}\nonumber\\
    &\quad{} + \chi_{([0, k/2,0,\ldots,0]; [2, k/2 + 1,0,\ldots,0])}
    + 2 \cdot\chi_{([0, k/2,0,\ldots,0]; [2, k/2 - 2, 1,0,\ldots,0])}\nonumber\\
   &\quad{}  + 2 \cdot\chi_{([0, k/2,0,\ldots,0]; [2, k/2 - 1,0,\ldots,0,1])}
    + 2 \cdot\chi_{([0, k/2,0,\ldots,0]; [0, k/2 , 1,0,\ldots,0,1])}\nonumber\\
    &\quad{} + \chi_{([0, k/2,0,\ldots,0]; [2, k/2 - 1, 1,0,\ldots,0])}
    + \chi_{([0, k/2,0,\ldots,0]; [2, k/2,0,\ldots,0,1])}+\mathcal{O}(q^2)\ ,\nonumber\\
    \mathcal{Z}_\text{vac}\cdot
(\chi^{(2)}_{+})_{{\ysm\yng(1,1)}} & = 
 \chi_{([0, k/2,0,\ldots,0];  [2, k/2,0,\ldots,0])} +
\chi_{([0, k/2,0,\ldots,0];  [2, k/2 - 1,0,\ldots,0])} \nonumber\\
& \quad {}+ \chi_{([0, k/2,0,\ldots,0]; [0, k/2,1,0,\ldots,0])} 
 +\chi_{([0, k/2,0,\ldots,0];  [0, k/2 + 1,0,\ldots,0,1])}  \nonumber\\
&  \quad {}+\chi_{([0, k/2,0,\ldots,0]; [2, k/2 + 1,0,\ldots,0])} 
+\chi_{([0, k/2,0,\ldots,0];  [2, k/2 - 2,0,\ldots,0])}\nonumber\\
&  \quad {} +\chi_{([0, k/2,0,\ldots,0]; [0, k/2,0,1,0,\ldots,0])}  
+\chi_{([0, k/2,0,\ldots,0]; [0, k/2 + 1,0,\ldots,0,1,0])}  
  \nonumber\\
&  \quad {}+2 \cdot\chi_{([0, k/2,0,\ldots,0];  [0, k/2,1,0,\ldots,0,1])}   \nonumber\\
&  \quad {} +\chi_{([0, k/2,0,\ldots,0];  [2, k/2 - 1,1,0,\ldots,0])} 
+\chi_{([0, k/2,0,\ldots,0];  [2, k/2 ,0,\ldots,0,1])} \nonumber\\ 
& \quad {}  +2\cdot \chi_{([0, k/2,0,\ldots,0]; [2, k/2 - 2,1,0,\ldots,0])}
+2 \cdot\chi_{([0, k/2,0,\ldots,0]; [2, k/2 - 1,0,\ldots,0,1])}
\nonumber\\
& \quad {} +\mathcal{O}(q^2)\ ,\nonumber \\
\mathcal{Z}_\text{vac}\cdot
(\chi^{(2)}_{-})_{{\ysm\yng(2)}} & = 
\chi_{([0, k/2,0,\ldots,0]; [2, k/2 ,0,\ldots,0])} 
+\chi_{([0, k/2,0,\ldots,0];  [2, k/2 - 1,0,\ldots,0])} \nonumber\\
& \quad {}+ \chi_{([0, k/2,0,\ldots,0]; [0, k/2 - 1,1,0,\ldots,0])} 
+ \chi_{([0, k/2,0,\ldots,0]; [0, k/2,0,\ldots,0,1])} \nonumber\\  
 & \quad {}+\chi_{([0, k/2,0,\ldots,0]; [2, k/2 +1,0,\ldots,0])} 
 + \chi_{([0, k/2,0,\ldots,0]; [2, k/2 - 2,0,\ldots,0])}  \nonumber\\
& \quad {}+ \chi_{([0, k/2,0,\ldots,0]; [0, k/2 - 1,0,1,0,\ldots,0])}  
+ \chi_{([0, k/2,0,\ldots,0]; [0, k/2 ,0,\ldots,0,1,0])} \nonumber\\
& \quad {}+ 2 \cdot\chi_{([0, k/2,0,\ldots,0]; [2, k/2 - 1,1,0,\ldots,0])}  
+ 2 \cdot\chi_{([0, k/2,0,\ldots,0];  [2, k/2,0,\ldots,0,1])}  \nonumber\\
& \quad {}+ 2 \cdot\chi_{([0, k/2,0,\ldots,0]; [0, k/2 - 1,1,0,\ldots,0,1])} \nonumber\\ 
& \quad {}+ \chi_{([0, k/2,0,\ldots,0]; [2, k/2 - 2,1,0,\ldots,0])}  
+ \chi_{([0, k/2,0,\ldots,0]; [2, k/2 - 1,0,\ldots,0,1])} \nonumber\\
& \quad {}+\mathcal{O}(q^2)\ ,\nonumber\\
\mathcal{Z}_\text{vac}\cdot (\chi^{(2)}_{-})_{{\ysm\yng(1,1)}} & = 
\chi_{([0, k/2,0,\ldots,0]; [0, k/2,0,\ldots,0])}
 +\chi_{([0, k/2,0,\ldots,0]; [2, k/2,0,\ldots,0])} \nonumber\\
 &\quad{} +\chi_{([0, k/2,0,\ldots,0]; [0, k/2-1,1,0,\ldots,0])}  
 +\chi_{([0, k/2,0,\ldots,0]; [0, k/2,0,\ldots,0,1])}  \nonumber\\ 
 &\quad{}+\chi_{([0, k/2,0,\ldots,0];[2, k/2-2,0,\ldots,0])} 
 +2 \cdot\chi_{([0, k/2,0,\ldots,0]; [2, k/2-1,1,0,\ldots,0])}  \nonumber\\
 &\quad{}+2 \cdot\chi_{([0, k/2,0,\ldots,0]; [2, k/2,0,\ldots,0,1])}  
 +2 \cdot\chi_{([0, k/2,0,\ldots,0]; [0, k/2-1,1,0,\ldots,0,1])} \nonumber\\ 
 &\quad{}+\chi_{([0, k/2,0,\ldots,0]; [2, k/2-2,1,0,\ldots,0])}
 +\chi_{([0, k/2,0,\ldots,0]; [2, k/2-1,0,\ldots,0,1])} +\mathcal{O}(q^2)\ ,\nonumber\\
\mathcal{Z}_\text{vac}\cdot \chi^{(2)}_{+}\chi^{(2)}_{-} & = \chi_{([0, k/2,0,\ldots,0]; [1, k/2,0,\ldots,0])}\nonumber\\
 &\quad{}+2 \cdot\chi_{([0, k/2,0,\ldots,0]; [1, k/2-1,1,0,\ldots,0])}
 +2 \cdot\chi_{([0, k/2,0,\ldots,0]; [1, k/2,0,\ldots,0,1])}\nonumber\\
 &\quad{}+\chi_{([0, k/2,0,\ldots,0];[1, k/2-1,0,\ldots,0])}
 +\chi_{([0, k/2,0,\ldots,0]; [1, k/2+1,0,\ldots,0])}\nonumber\\
 &\quad{}+2 \cdot\chi_{([0, k/2,0,\ldots,0]; [1, k/2-2,1,0,\ldots,0])}
 +2 \cdot\chi_{([0, k/2,0,\ldots,0]; [1, k/2-1,0,\ldots,0,1])}\nonumber\\
 &\quad{}+4 \cdot\chi_{([0, k/2,0,\ldots,0]; [1, k/2-1,1,0,\ldots,0,1])} \nonumber\\ 
 &\quad{}+\chi_{([0, k/2,0,\ldots,0]; [1, k/2-1,0,1,0,\ldots,0])}
+\chi_{([0, k/2,0,\ldots,0]; [1, k/2,0,\ldots,0,1,0])}\nonumber\\
 &\quad{}+2 \cdot\chi_{([0, k/2,0,\ldots,0];[1, k/2,1,0,\ldots,0])}
 +2 \cdot\chi_{([0, k/2,0,\ldots,0]; [1, k/2+1,0,\ldots,0,1])}\nonumber\\
 &\quad{}+2 \cdot\chi_{([0, k/2,0,\ldots,0]; [3, k/2,0,\ldots,0])}
 +2 \cdot\chi_{([0, k/2,0,\ldots,0]; [3, k/2-1,0,\ldots,0])}\nonumber\\
&\quad{} +2 \cdot\chi_{([0, k/2,0,\ldots,0]; [3, k/2-2,0,\ldots,0])}+\mathcal{O}(q^2)\ .
\end{align}

The systematics of the decompositions are analogous
to the 2-cycle twist case, see the  discussion following eq.~(\ref{3.14}) above,
but are somewhat more complicated. 
Each box appended to the first two rows of $\Lambda_-$ 
of the ground state
\eqref{eq:22_gs} has odd parity under one of the two $\mathbb{Z}_2$'s.
As a consequence, the states that appear in the 
mixed sector $\chi^{(2)}_{+}\chi^{(2)}_{-} $
are precisely those that have an odd number of them, i.e., for which $l_1$ is odd. 
Conversely, the other four representations
contain the states with $l_1$ even, but the selection rules among them are more subtle,
and indeed, the same coset representation can 
appear in different $D_8$ decompositions. 
For example, the lowest state in the representation 
\begin{equation}\label{testrep}
\Lambda_- = [2,k/2,0,\ldots,0]
\end{equation}
can be constructed as an excitation of the twisted sector 
ground state with a fermionic zero-mode and 
a bosonic $(-\tfrac{1}{2})$-mode involving the same twisted coordinate. Then the 
state has $(+,+)$ charge under $\mathbb{Z}_2^2$,
and we can either symmetrise or anti-symmetrise it with respect
to the $S_2$ factor. That is why this state appears
both in $(\chi^{(2)}_{-})_{{\ysm\yng(2)}}$ and
in $(\chi^{(2)}_{-})_{{\ysm\yng(1,1)}}$. But we can also
construct the lowest state of (\ref{testrep}) by exciting the twisted sector ground state
with a fermionic zero-mode from one twisted coordinate, 
and a bosonic $(-\tfrac{1}{2})$-mode
from the other,  and symmetrise with respect 
to $S_2$.\footnote{The antisymmetric combination is 
actually a supersymmetric descendant of the excitation by the two
fermionic zero-modes and is therefore part of
$([0,k/2,0,\ldots,0];[0,k/2+1,0,\ldots,0])$.}
In this case the charge is $(-,-)$ under $\mathbb{Z}_2^2$, 
and the state is even under the $S_2$; 
thus the representation (\ref{testrep}) also appears in the decomposition of 
$(\chi^{(2)}_{+})_{{\ysm\yng(1,1)}}$.

\subsection{Sectors of arbitrary twist}
\label{sec:gen_twist}

While the detailed description of the decompositions 
into $s{\cal W}_\infty$ characters becomes
more and more cumbersome, some aspects of the 
twisted sector can be described quite generally. 
In particular, the partition function of any 
twisted sector can be written in `multiparticle' form, generalising
eq.~(\ref{eq:structure22}).\footnote{This observation 
was recently also made in \cite{Jevicki:2015irq}.}
Let us first explain this for the twisted sectors $(2)^n$ corresponding to 
multiple $2$-cycle twists. By the DMVV formula \eqref{eq:gen_fct_orb_RR},  the 
generating function for this part of the partition function in the R-R sector equals 
\begin{align}
\label{eq:gen_n2cycles}
  \sum_{N=0}^{\infty}p^N Z^{(2)^n}_\text{R}(S^N \mathbb{T}^2)
&=\left. \sideset{}{'}\prod_{\Delta,\bar\Delta,\ell,\bar\ell}
 \frac{1}{\left(1-(-1)^{\ell+\bar\ell+1}p^2\,q^\frac{\Delta}{2}\bar{q}^\frac{\bar\Delta}{2}
 y^{\ell}\bar{y}^{\bar\ell}\right)^{c(\Delta,\bar\Delta,\ell,\bar\ell)}} \,
\right|_{p^{2n}}  \nonumber\\
&\qquad{}\times p^{2n}\prod_{\Delta,\bar\Delta,\ell,\bar\ell}
 \frac{1}{\left(1-(-1)^{\ell+\bar\ell+1}p\,q^{\Delta}\bar{q}^{\bar\Delta}
 y^{\ell}\bar{y}^{\bar\ell}\right)^{c(\Delta,\bar\Delta,\ell,\bar\ell)}}\ .
\end{align}
We recognise the second factor as the untwisted partition function
of $S_{N-2n}$, which is indistinguishable from the untwisted partition
function of $S_N$ as $N\to\infty$.
The first factor, on the other hand, can be expressed in terms of sums of squares 
of all possible symmetrisations of the elementary characters 
$\chi^{(2)}_{\pm}$. To see this, let us write
\begin{multline}
\left. \sideset{}{'}\prod_{\Delta,\bar\Delta,\ell,\bar\ell}
 \frac{1}{\left(1-(-1)^{\ell+\bar\ell+1}p^2\,q^\frac{\Delta}{2}\bar{q}^\frac{\bar\Delta}{2}
 y^{\ell}\bar{y}^{\bar\ell}\right)^{c(\Delta,\bar\Delta,\ell,\bar\ell)}}\, 
\right|_{p^{2n}}={}\\
\begin{aligned}
&=\left. \exp \left[-\!\! \sideset{}{'}
\sum_{\Delta,\bar\Delta,\ell,\bar\ell}
c(\Delta,\bar\Delta,\ell,\bar\ell)\log\!\left(1-(-1)^{\ell+\bar\ell+1}p^2\,q^\frac{\Delta}{2}\bar{q}^\frac{\bar\Delta}{2}
 y^{\ell}\bar{y}^{\bar\ell}\right)\right] \right|_{p^{2n}}\nonumber\\
&=\left. \exp \left[\sideset{}{'}
\sum_{\Delta,\bar\Delta,\ell,\bar\ell}
c(\Delta,\bar\Delta,\ell,\bar\ell)\sum_{k=1}^{\infty}\frac{p^{2k}}{k}(-1)^{k(\ell+\bar\ell+1)}q^{\frac{k\Delta}{2}}\bar{q}^{\frac{k\bar\Delta}{2}}
 y^{k\ell}\bar{y}^{k\bar\ell}\right] \right|_{p^{2n}}\ .
\end{aligned}
\end{multline}
Changing the order of summation and flowing to the NS-NS sector, this becomes
\begin{multline}
\hspace*{-0.3cm} \left. \exp \left[\sum_{k=1}^{\infty}\frac{p^{2k}}{k}
\sideset{}{'}\sum_{\Delta,\bar\Delta,\ell,\bar\ell}
c(\Delta,\bar\Delta,\ell,\bar\ell)(-1)^{k(\ell+\bar\ell+1)}q^{\frac{k}{2}(\Delta+\ell+1)}\bar{q}^{\frac{k}{2}(\bar\Delta+\bar\ell+1)}
 y^{k(\ell+1)}\bar{y}^{k(\bar\ell+1)}\right] \right|_{p^{2n}} \\
\begin{aligned}
{} &=\left. \exp \left[ \sum_{k=1}^{\infty}\frac{p^{2k}}{k}
\left(\lvert\mathcal{F}^{k-1}\chi^{(2)}_{+}(q^k,y^k)\rvert^2
+\lvert\mathcal{F}^{k-1}\chi^{(2)}_{-}(q^k,y^k)|^2\right)\right]\right|_{p^{2n}}\nonumber\\
{}  &=\sum_{m=0}^n \!\left.\exp \!\!\left[\sum_{k=1}^{m}\frac{p^{2k}}{k}
\lvert\mathcal{F}^{k-1}\chi^{(2)}_{+}(q^k,y^k)\rvert^2
\right] \!\right|_{p^{2m}}\!\!\!\cdot
\left.\exp \!\!\left[ \sum_{k=1}^{n-m}\frac{p^{2k}}{k}
\lvert\mathcal{F}^{k-1}\chi^{(2)}_{-}(q^k,y^k)\rvert^2
\right]\!\right|_{p^{2(n-m)}}\!  .
\end{aligned}
\end{multline}
Next we note that 
\begin{align}
\left.\exp \left[ \sum_{j=1}^{m}\frac{p^{2j}}{j}
\lvert\mathcal{F}^{j-1}\chi(q^j,y^j)\rvert^2 \right] \right|_{p^{2m}}
&=\sum_{\substack{k_1,\ldots,k_m\geq0 \\ \sum_j j k_j = m}}
\frac{1}{\prod_{i=1}^m i^{k_i} k_i!}\prod_{j=1}^m 
\lvert\mathcal{F}^{j-1}\chi(q^j,y^j)\rvert^{2k_j}\nonumber\\
&=\frac{1}{m!} \sum_{\rho \in S_m} \prod_{j=1}^m 
\lvert\mathcal{F}^{j-1}\chi(q^j,y^j)\rvert^{2a_j(\rho)} \nonumber\\
&=\sum_{\Lambda \in Y_m} \lvert\chi_{\Lambda}(q,y)\rvert^2\ .
\end{align}
In the second equality, we have used that $m!/\!\prod_{i=1}^m i^{k_i} k_i!$ is the
number of elements in the conjugacy class 
$C_{k_1,\ldots,k_m}$ of $S_m$, which consist
of $k_i$ cycles of length $i$. On the other hand, the last equality follows from 
\eqref{eq:char_symm} and the column orthogonality of $S_m$ characters,
\begin{equation}
  \sum_{\Lambda \in Y_m}
\left(\chi_m^{\Lambda}(\rho)\right)^2=\frac{m!}{|C_{\rho}|} 
\qquad \text{for any }\rho \in S_m\ .
\end{equation}
Here the sum is over all Young diagrams of $m$ boxes 
or all irreducible representations of $S_m$.
It follows that
\begin{equation}
 Z^{(2)^n}(S^N \mathbb{T}^2)=Z^\text{(U)}(S^{N-2n}\mathbb{T}^2)\cdot\sum_{k=0}^n
\sum_{\Lambda_1 \in Y_k}\lvert(\chi^{(2)}_{+})_{\Lambda_1}(q,y)\rvert^2 
\sum_{\Lambda_2 \in Y_{n-k}}\lvert(\chi^{(2)}_{-})_{\Lambda_2}(q,y)\rvert^2\ ,
\end{equation}
thus generalising (\ref{eq:structure22}) to the case $n>2$. 
\smallskip

So far we have only dealt with multiple $2$-cycles, 
but the analysis is fairly analogous for 
the twist $(m)^n$ consisting of $n$ non-overlapping $m$-cycles. The analogue of 
eq.~\eqref{eq:gen_n2cycles} for $m\geq 2$ is now
\begin{align}
\label{eq:gen_nmcycles}
\sum_{N=0}^{\infty}p^N Z^{(m)^n}_\text{R}(S^N \mathbb{T}^2)
&= \left.{\prod_{\substack{\Delta,\bar\Delta,\ell,\bar\ell\\
m|(\Delta-\bar{\Delta})}}}
 \frac{1}{\left(1-(-1)^{\ell+\bar\ell+1}p^m\,
 q^{\frac{\Delta}{m}}\bar{q}^{\frac{\bar\Delta}{m}}
 y^{\ell}\bar{y}^{\bar\ell}\right)^{c(\Delta,\bar\Delta,\ell,\bar\ell)}}\,
\right|_{p^{mn}} 
\nonumber\\
&\qquad{}\times p^{mn}\prod_{\Delta,\bar\Delta,\ell,\bar\ell}
 \frac{1}{\left(1-(-1)^{\ell+\bar\ell+1}p\,q^{\Delta}
\bar{q}^{\bar\Delta}
 y^{\ell}\bar{y}^{\bar\ell}\right)^{c(\Delta,\bar\Delta,\ell,\bar\ell)}}\ .
\end{align}
The analysis goes through essentially unmodified, and we find that we can express the 
partition function of this sector in terms of the elementary characters
\begin{equation}\label{building}
 \chi^{(m)}_i(q,y) = \sum_{\substack{\Delta,\ell\\\Delta \equiv i \,(\text{mod } m)}}
\!\!\!\!\!\abs{c(\Delta,\ell)}\,
q^{\frac{\Delta}{m}+\frac{\ell}{2}+\frac{m}{4}}\,y^{\ell+\frac{m}{2}}
\quad \text{ for } i=1,\ldots,m
\end{equation}
as  
\begin{equation}
  Z^{(m)^n}(S^N \mathbb{T}^2)=Z^\text{(U)}(S^{N-mn}\mathbb{T}^2)\cdot
\sum_{\substack{k_1,\ldots,k_m\geq0\\\sum_j k_j = n}}\,
\prod_{i=1}^m \sum_{\Lambda \in Y_{k_i}}
\lvert( \chi^{(m)}_i)_{\Lambda}(q,y)\rvert^2 \ .
\end{equation}
In particular, in the sector whose twist is just one cycle of length $m$,
we have $n=1$ and thus
\begin{equation}
  Z^{(m)}(S^N \mathbb{T}^2)=Z^\text{(U)}(S^{N-m}\mathbb{T}^2)\cdot
\sum_{i=1}^m
\lvert\chi^{(m)}_i(q,y)\rvert^2 \ .
\end{equation}

It remains to combine these statements to cover the general case of a twist 
with cycle structure $(1)^{N_1}(2)^{N_2}\cdots(n)^{N_n}$, i.e.,
$N_i$ cycles of length $i$ for $i=1,\ldots,n$, where $\sum_i iN_i=N$. 
By the DMVV formula \eqref{eq:gen_fct_orb_RR},
the R-R partition function factorises into $n$ components pertaining
to the different cycle lengths
\begin{equation}
Z_\text{R}^{(1)^{N_1}\cdots(n)^{N_n}}(S^N \mathbb{T}^2)
= \left. \prod_{m=1}^n {\prod_{\substack{\Delta,\bar\Delta,\ell,\bar\ell\\
m|(\Delta-\bar{\Delta})}}}
 \frac{1}{\left(1-(-1)^{\ell+\bar\ell+1}p^m\,
 q^\frac{\Delta}{m}\bar{q}^\frac{\bar\Delta}{m}
 y^{\ell}\bar{y}^{\bar\ell}\right)^{c(\Delta,\bar\Delta,\ell,\bar\ell)}}
\, \right|_{p^{mN_m}}  \ ,
\end{equation}
and correspondingly for the NS-NS sector.
Plugging in our results from above, we obtain
\begin{align}
Z^{(1)^{N_1}\cdots(n)^{N_n}}(S^N \mathbb{T}^2)
&=\prod_{m=1}^n \sum_{\substack{k_1,\ldots,k_m\geq0\\\sum_j k_j = N_m}}\,
\prod_{i=1}^m \sum_{\Lambda \in Y_{k_i}}
\lvert( \chi^{(m)}_i)_{\Lambda}(q,y)\rvert^2\nonumber\\
&= Z^\text{(U)}(S^{N_1}\mathbb{T}^2)\cdot\prod_{m=2}^n
\sum_{\substack{k_1,\ldots,k_m\geq0\\\sum_j k_j = N_m}}\,
\prod_{i=1}^m \sum_{\Lambda \in Y_{k_i}}
\lvert( \chi^{(m)}_i)_{\Lambda}(q,y)\rvert^2 \ . \label{twistmulti}
\end{align}
Thus we can think of the entire twisted sector as consisting 
of the `multiparticle' contributions
of the fundamental building blocks (\ref{building}). 
\medskip

As was already alluded to before, essentially the 
same techniques also allow us to prove the identity 
eq.~\eqref{eq:part_untw} for the untwisted sector partition function. Since 
$\chi^{(1)}_{1}(q,y)=Z^\text{(chiral)}_\text{NS}(\mathbb{T}^2)(q,y)=1+\chi_1(q,y)$ and
\begin{equation}
\label{eq:vac_prod}
 \mathcal{Z}_\text{vac}'(q,y)=\prod_{\substack{(\Delta,\ell)\\\neq(0,-\frac{1}{2})}}
 \frac{1}{\left(1-q^{\Delta+\frac{\ell}{2}+\frac{1}{4}}
 (-y)^{\ell+\frac{1}{2}}\right)^{c(\Delta,\ell)}}\ ,
\end{equation}
we get, for $N\to\infty$,
\begin{align}
 Z^\text{(U)}(S^N \mathbb{T}^2)(q,\bar{q},y,\bar{y})&=
\prod_{\substack{(\Delta,\bar\Delta,\ell,\bar\ell)\nonumber\\
\neq(0,0,-\frac{1}{2},-\frac{1}{2})}}
 \frac{1}{\left(1-q^{\Delta+\frac{\ell}{2}+\frac{1}{4}}
\bar{q}^{\bar\Delta+\frac{\bar\ell}{2}+\frac{1}{4}}
 (-y)^{\ell+\frac{1}{2}}(-\bar{y})^{\bar\ell+
 \frac{1}{2}}\right)^{c(\Delta,\bar\Delta,\ell,\bar\ell)}}\nonumber\\
&=|\mathcal{Z}_\text{vac}'(q,y)|^2 \cdot
\exp \sum_{k=1}^{\infty}\frac{1}{k}
\lvert\mathcal{F}^{k-1}\chi_{1}(q^k,y^k)\rvert^2\nonumber\\
&=|\mathcal{Z}_\text{vac}'(q,y)|^2 
\sum_{m=0}^\infty \sum_{\Lambda \in Y_m}
|\chi_{\Lambda}(q,y)|^2\ ,
\end{align}
which reproduces \eqref{eq:part_untw} upon dividing 
by $Z_\text{NS}(\mathbb{T}^2)$, see eq.~\eqref{2.15}.

\subsection{Twisted representations of the wedge algebra}\label{sec:3.4}

Given the multiparticle structure of the entire twisted sector, see eq.~(\ref{twistmulti}), 
it only remains to understand the structure of the building blocks 
$\chi^{(m)}_i$ (that account for the individual `particles'). These wedge characters
count states that sit in representations of the 
wedge subalgebra $\mathrm{shs}[\mu]$ of $s\mathcal{W}_{\infty}[\mu]$.
In this section we undertake first steps to understand 
the structure of these higher spin representations.
This should shed light on the `particle' structure of the 
stringy extension of the higher spin theory;
in \cite{Gaberdiel:2015wpo} the relevant analysis 
was done for the bosonic toy model consisting
of a single boson, here we explain the ${\cal N}=2$ generalisation. 

As was explained at the beginning of this chapter, 
the $m$-cycle twisted sector is generated by
complex fermions and bosons of twist $\xi_i=\frac{i}{m}$, 
where $i=1,\ldots, m$. Since the $s{\cal W}_\infty$ generators
are neutral bilinears in the currents (and since their mode 
numbers continue to be integers or half-integers
depending on the statistics), the contribution
coming from the individual twisted (complex) bosons 
and fermions decouple from one another,
and we can think of the representation as consisting 
of an $m$-fold tensor product of the individual
twist $\xi_i$ contributions. Apart from one untwisted 
component corresponding to $i=m$ --- this does not contribute 
to the wedge  character --- the other $(m-1)$ components 
all lead to  representations whose wedge character
is of the form (see also \cite{Gaberdiel:2015wpo}) 
\begin{align}
\chi_{\xi} (q,y) = q^h\left.\prod_{n=1}^{\infty}
\frac{( 1+zyq^{n-\frac{1}{2}-\xi})( 1+z^{-1}y^{-1}q^{n-\frac{1}{2}+\xi}) }
{(1-zq^{n-\xi })(1-z^{-1}q^{n-1+\xi})}\right|_{z^p}\ .
\end{align}
Here we have assumed that $0<\xi<\frac{1}{2}$, 
and $z$ keeps track of the twist, i.e., 
the terms with a given power of $z^p$ pick up the same phase under the 
cyclic group $\mathbb{Z}_m$ in the centraliser. In the following, we shall concentrate on the 
$z^0$ case, for which the states transforms trivially under $\mathbb{Z}_m$.
The $q$-expansion of this character is 
\begin{align}\label{eq:twist_character}
\chi_{\xi} (q,y) = q^h\left(1 +y  q^\frac{1}{2}	+ 2 q 
+ (3y + y^{-1}) q^\frac{3}{2} + (y^2 + 6 ) q^2 
+ (8y+3y^{-1}) q^\frac{5}{2} + \ldots 	 \right)\ .
\end{align}
For $\xi<\frac{1}{2}<1$ there is a similar answer 
where $y$ is replaced by $y\mapsto y^{-1}$; the case $\xi=\frac{1}{2}$ is 
a bit special since there are then fermionic zero modes. 

Each such representation has a single descendant 
at level $1/2$, and is therefore a special 
case of what one may like to call a `level-1/2 representation', 
compare the terminology of 
\cite{Gaberdiel:2015wpo}. Thus we can learn about the 
structure of the twisted sector by studying 
general level-1/2 representations, and this is what we
shall be doing in the following.
\smallskip

Suppose $\phi$ is the ground state of a level-1/2 representation. 
Let us assume for definiteness 
that $\phi$ is annihilated by $G^-_{-1/2}$ (rather than $G^+_{-1/2}$), i.e., 
\begin{align}
G^-_{-1/2} \phi =0\ ,
\end{align}
as well as by all the other negative charge fermionic spin $s$ supercharges, i.e.,
\begin{align}\label{W-N}
W^{s-}_{-1/2}\phi   =0  \qquad \hbox{for $s=2,3,\ldots$} \ .
\end{align}
(This is the situation that is relevant 
for (\ref{eq:twist_character}); the conjugate solution arises for 
$\frac{1}{2} < \xi < 1$.) Here we have denoted the 
generators of the spin $s$ multiplet by (see e.g., \cite{Candu:2012tr})
\be
W^{s0} \ , \qquad W^{s\pm}  \ , \qquad W^{s1} 
\ee
of spin $s$, $s+\frac{1}{2}$, and $s+1$, respectively. 
The corresponding modes then transform in a representation of the
superconformal algebra 
\begin{align}
[G^\pm_r,W^{s0}_n] & = \mp W^{s\pm}_{r+n} \nonumber\\
\{ G^\pm_r, W^{s\pm}_r \} & = 0 \nonumber\\
\{G^\pm_r, W^{s\mp}_r \} & = \pm \bigl( (2s-1)r - t \bigr)\, W^{s0}_{r+t} + 2 W^{s1}_{r+t} \nonumber\\
{}[G^\pm_r,W^{s1}_n] & =  \bigl( sr - \tfrac{1}{2}\, n\bigr) \, W^{s\pm}_{r+n} \ . 
\end{align}

Let us denote  the eigenvalues of the zero-modes 
$W^{s0}_0$ and $W^{s1}_0$ on the ground state $\phi$
by $w^{s0}$ and $w^{s1}$, respectively. Then it follows from (\ref{W-N}) that 
\be
0 = G^+_{1/2} \, W^{s-}_{-1/2} \, \phi = \bigl( s w^{s0} + 2 w^{s1} \bigr) \, \phi
\ee
and hence
\be\label{s1s0rel}
w^{s1} = -\frac{1}{2} s w^{s0} \ . 
\ee
Note that for $s=1$ this reduces to the familiar chiral primary condition, 
namely that $h=-\frac{1}{2} q$, where
$q=w^{10}$ is the ${\rm U}(1)$ charge with respect to the 
spin $1$ field in the ${\cal N}=2$ supermultiplet,  and 
$h=w^{11}$ is the conformal dimension.

The other condition that follows from the level-1/2 condition 
is that all the states generated by the $W^{s+}_{-1/2}$ modes 
from the ground state are proportional to $ G^{+}_{-1/2}  \phi$, i.e.,
\begin{align}\label{other-null-vecs}
W^{s+}_{-1/2} \phi  = \alpha(s) \, G^{+}_{-1/2}  \phi \ .
\end{align}
Applying $G^{-}_{1/2}$ to this relation and using the 
above commutation relations, we find that
\begin{align}
\alpha(s) = - \frac{sw^{s0}}{2h}\ ,
\end{align}
where we have used \eqref{s1s0rel}. 

In order to obtain a relation between the different 
quantum numbers $\alpha(s)$, we finally apply the $W^{20}_0$ mode to 
both sides of eq.~\eqref{other-null-vecs}. For example, 
for the case where $s=2$ and using the $[W^{20}_m,W^{2+}_r]$ 
commutation relation, we conclude that
\begin{align}\label{alpha32}
\alpha(3) &= - \frac{8q_3(5\nu^2 -8\sqrt{3}\nu \alpha(2)-15(8\alpha(2)^2 + 3))}{9(\nu-5)}\ ,
\end{align}
where $\nu = 2\mu-1$ and $q_3$ is a normalisation constant of $W^{30}$.
This determines $\alpha(3)$ as a function of $\alpha(2)$. 
Continuing in this manner, we obtain a recursion relation for all $\alpha(s)$. 
This shows that all higher quantum numbers $w^{s0}$ and 
$w^{s1}$ are recursively determined.
Thus the assumption that there is a single state at level $1/2$ implies that 
the most general level-1/2 representation is 
characterised by only two quantum numbers
\begin{align}\label{paras}
h \equiv w^{11} \quad \text{and} \quad \alpha(2)\equiv -\frac{w^{20}}{h}\ .
\end{align}

\subsection{A relation between the parameters}\label{sec:3.5}

As in the bosonic analysis of \cite{Gaberdiel:2015wpo}, 
it seems that the actual $\xi$-twisted 
representation is a special type of level-1/2 representation, 
and has in fact one fewer state
at level 3/2 than a generic level-1/2 
representation.\footnote{This will be discussed in more detail
in \cite{DattaGaberdiel}.}  One should therefore 
expect that it is characterised by a special relation 
between the two eigenvalues in (\ref{paras}). In order
to work out what this relation should be, we can use 
that the $\xi$-twisted representation
is described, in the coset language, 
by the large $k$ limit of the coset representation
$([\xi k,0,\ldots,0]; [\xi k,0,\ldots,0])$ \cite{Gaberdiel:2014vca}.  
In order to evaluate the 
eigenvalues of $T$ and $W^{20}$ on this coset state, 
we have worked out the form of the spin $2$ fields
in the coset; this is discussed, in some detail, 
in the appendix. With the notation of the appendix, in particular, 
\eqref{eq:T}, \eqref{eq:w20}, \eqref{eq:w20_coeff} 
and \eqref{eq:alphabeta}, we find that
in the (large $c$ and $\nu=-1$) 't~Hooft limit
\begin{align}\label{free-currents}
T&= T_{\rm b} + T_{\rm f}  + \frac{3}{2c}\, J^2\ ,\nn\\
W^{20} &=  \frac{1}{\sqrt{3}} \left( - T_{\rm b} +2\, T_{\rm f}  \right)\ .
\end{align}
The mode expansions of the stress tensor of a 
single free boson and fermion  are given by (the fermion has NS boundary conditions)
\begin{align}
(L_{\rm b})_m &= \sum_{n \in \mathbb{Z}} ^\infty  
\normord{ \bar\alpha_{m-n}  \alpha_n} \ , \nn \\
(L_{\rm f})_m &= \frac{1}{2}
\sum_{r \in \mathbb{Z}+1/2} ^\infty (2r-m)\, \normord{\bar{\psi}_{m-r} \psi_r }\ .
\end{align}
Here the bosonic and fermionic modes satisfy the usual commutation relations
\begin{align}
[\alpha_m,\alpha_n]=0=[\bar\alpha_m,\bar\alpha_n] \ , 
\qquad  &[\alpha_m, \bar\alpha_n] = m \delta_{m,-n} \ ,\nonumber\\
\lbrace \psi_r , \psi_s  \rbrace = 0 = \lbrace \bar\psi_r ,  \bar\psi_s  \rbrace \ ,
 \qquad  &\lbrace \psi_r , \bar{\psi}_s   \rbrace = \delta_{r,-s}\ .
\end{align}
In the $\xi$-twisted sector, the boson and fermion 
mode numbers get shifted, and the zero mode of the stress tensor
picks up a normal-ordering contribution
\begin{align}
(L_{\rm b})_0 &= \sum_{r\in\mathbb{Z}+\xi} \normord{\bar\alpha_{-r} \alpha_{r}} \ 
+ \ \frac{1}{2}\, \xi\, (1-\xi) \ ,  \nn \\
(L_{\rm f})_0 &=  \sum_{s\in\mathbb{Z}+\frac{1}{2}+\xi}\!\!\! s \,
\normord{ \bar\psi_{-s} \psi_s} \ + \ \frac{\xi^2}{2}  \ .
\end{align}
For large $c$ we then find for the 
eigenvalues of $T_0$ and $W^{20}_0$ 
\begin{align}
h= \frac{\xi}{2} \ , \qquad w^{20}= \frac{1}{2\sqrt{3}} \, \xi (3\xi-1) \ .
\end{align}
Eliminating $\xi$ from the above equations yields
\begin{align}\label{eq:alpha_root1}
\alpha(2) = - \frac{w^{20}}{h} = - \frac{1}{\sqrt{3}}\, (3\xi-1) = 
-\,\frac{6h-1}{\sqrt{3}}  \ .
\end{align}
This is therefore the additional relation which characterises 
the special level-1/2 representations
that arise in the twisted sector.

\section{Conclusions}

In this paper we have analysed the embedding of the ${\cal N}=2$ cosets
that appear in the duality with the ${\cal N}=2$ supersymmetric higher spin theory
on AdS$_3$ into the symmetric orbifold of $\mathbb{T}^2$. This is the ${\cal N}=2$
analogue of the ${\cal N}=4$ construction of \cite{Gaberdiel:2014cha} where
the relevant symmetric orbifold is known to be dual to string theory on 
AdS$_3\times {\rm S}^3\times \mathbb{T}^4$. It is therefore tempting to believe,
in particular given the recent discussions of 
\cite{Hartman:2014oaa,Haehl:2014yla,Belin:2014fna,Belin:2015hwa}, 
that also the  symmetric orbifold of $\mathbb{T}^2$ should be dual to some string
theory on AdS$_3$. For example, a candidate background 
could be the (warped) product of the form 
\be
{\rm AdS}_3 \times {\rm S}^3 \times 
\bigl( \mathbb{T}^2 \times \mathbb{T}^2 \bigr) / S_2 \ , 
\ee
where the $S_2$ exchanges the two $\mathbb{T}^2$'s --- this is not too dissimilar
to the background with $(4,2)$ superconformal symmetry 
found in \cite{Donos:2014eua}.\footnote{As far as we are aware, no supergravity background
with $(2,2)$ superconformal symmetry is explicitly known, although such backgrounds probably
exist. We thank Jerome Gauntlett for a correspondence about this point.} Alternatively,
one may want to replace the symmetric orbifold with an orbifold with respect to 
a smaller group, e.g., 
\be
\bigl( \mathbb{T}^2 \bigr)^{2N} / ( S_2^N \rtimes S_N) \ ,
\ee
where $S_2^N\rtimes S_N$ is the so-called wreath product, 
i.e., the semidirect product which contains
$S_2^N$ as a subgroup on which $S_N$ acts in the obvious manner. 
Since the wreath product is a subgroup of the full 
permutation group, $S_2^N \rtimes S_N \subset S_{2N}$, 
the corresponding conformal field theory defines an even 
further extension of the symmetric orbifold we 
have considered above. In particular, it therefore 
contains the ${\cal N}=2$ Kazama-Suzuki models that are dual to the 
higher spin theory on AdS$_3$.

Part of the motivation for studying the ${\cal N}=2$ version of the duality is that
the Kazama-Suzuki models that appear in the dual of the higher spin theory 
\cite{Creutzig:2011fe,Candu:2012jq} correspond to 
the special family of ${\cal N}=2$ cosets 
\begin{equation}
\label{eq:ext_coset}
\frac{\mathfrak{su}(N+M)^{(1)}_{k+N+M}}{\mathfrak{su}(N)^{(1)}_{k+N+M} \oplus 
\mathfrak{su}(M)^{(1)}_{k+N+M} \oplus \mathfrak{u}(1)^{(1)}_\kappa}\  
\end{equation}
with $M=1$. The cosets therefore allow for a `matrix-like' 
extension ($M>1$), similar to what was 
considered for the case of AdS$_4$ in \cite{Chang:2012kt}, 
and it would be very interesting
to understand the correct AdS$_3$ description of this construction. 
First steps in this direction
were already undertaken in \cite{Creutzig:2014ula}, 
but it would be very instructive to repeat
the analysis of the present paper for these more 
general cosets, and see how the results
fit together with permutation orbifold theories 
that may have a fairly direct stringy interpretation. 

We have also analysed the representations 
of the higher spin algebra that arise
in the twisted sector; a good understanding of 
these representations will be key
for characterising the stringy extension from a 
higher spin viewpoint. While
some aspects of the description were 
rather similar to the bosonic analysis of 
\cite{Gaberdiel:2015wpo}, it seems 
that there are also interesting and subtle
differences; these will be explored 
further in \cite{DattaGaberdiel}.

\section*{Acknowledgments}
We thank Shouvik Datta for many useful discussions about various aspects of this paper and an initial collaboration
about aspects of Sections~\ref{sec:3.4} and \ref{sec:3.5}. We also
thank Marco Baggio, Constantin Candu, Jerome Gauntlett, Rajesh Gopakumar, Christoph Keller, Wei Li, 
and Cheng Peng for helpful discussions. The work of 
M.K.\ was supported by a studentship from the Swiss National Science Foundation. This research was also
(partly) supported by the NCCR SwissMAP, funded by the Swiss National Science Foundation.

\appendix

\section{The coset analysis}
\label{app:coset}
In this appendix we explain in some detail
the construction of the spin $1$ and $2$ currents of the coset
\begin{equation}
\label{eq:app_coset}
\frac{\mathfrak{su}(N+1)^{(1)}_{k+N+1}}{\mathfrak{su}(N)^{(1)}_{k+N+1} \oplus \mathfrak{u}(1)^{(1)}_\kappa}\  
\end{equation}
with $\kappa=N(N+1)(N+k+1)$.
We will closely follow the analysis of \cite{Gaberdiel:2014yla}
in the $\mathcal{N}=4$ case and \cite{ZhuliHe2014}.
The numerator consists of $N(N+2)$ bosonic currents 
$\mathcal{J}^A$ and free fermions $\psi^A$
transforming in the adjoint representation of
$\mathfrak{su}(N+1)$.
Given a hermitian orthonormal basis $t_{ij}^A$ of
$\mathfrak{su}(N+1)$ satisfying
\begin{equation}
 [t^A,t^B]=if^{ABC}t^C
 \qquad \text{and} \qquad
 \Tr (t^A\,t^B) = \delta^{AB}\ ,
 \end{equation}
which we order in such a way that
$t^a$ for $a=1,\ldots,N^2-1$ form a hermitian
orthonormal basis of $\mathfrak{su}(N)$, 
the numerator fields satisfy the commutation relations
 \begin{align}
 [\mathcal{J}^A_m,\mathcal{J}^B_n]
 &=if^{ABC} \mathcal{J}^C_{m+n}+(k+N+1)\delta^{AB}\delta_{m,-n}\ ,\nonumber\\
  [\mathcal{J}^A_m,\psi^B_r]
 &=if^{ABC} \psi^C_{m+r}\ ,\nonumber\\
 \lbrace\psi^A_r,\psi^B_s\rbrace
 &=\delta^{AB}\delta_{r,-s}\ .
 \end{align}
Restricting the adjoint representation to the
 denominator subalgebra, it decomposes as
 \begin{equation}
 \mathfrak{su}(N+1)\to \mathfrak{su}(N)
 \oplus \mathfrak{u}(1) \oplus \mathbf{N}
 \oplus \mathbf{\bar{N}}\ .
 \end{equation}
 We can decouple the currents from the fermions
 by defining
 \begin{equation}
 J^A = \mathcal{J}^A +\frac{i}{2}f^{ABC}(\psi^B\psi^C)
\end{equation}
 in the numerator or 
  \begin{equation}
 \widetilde{J}^a = \mathcal{J}^a +\frac{i}{2}f^{abc}(\psi^b\psi^c)
\end{equation}
in the denominator, where again lower-case indices
from the beginning of the alphabet range from $1$
to $N^2-1$ only. These currents and the fermion bilinears 
give rise to the bosonic coset
\begin{equation}
\frac{\mathfrak{su}(N+1)_{k}\oplus \mathfrak{so}(2N)_1}
{\mathfrak{su}(N)_{k+1} \oplus \mathfrak{u}(1)_\kappa}  \ .
\end{equation}
From the $N(N+2)$ fermions in the numerator
we subtract the $N^2$ fermions in the denominator.
The $2N$ surviving fermions can be defined by
\begin{equation}
\psi^i = t^A_{N+1,i}\psi^A\ ,
\qquad
\bar{\psi}^i = t^A_{i,N+1}\psi^A\ ,
\end{equation}
satisfying
\begin{gather}
\lbrace\psi^i_r,\bar{\psi}^j_s\rbrace = \delta^{ij}\delta_{r,-s}\ ,\nonumber\\
\lbrace\psi^i_r,\psi^j_s\rbrace = \lbrace\bar{\psi}^i_r,\bar{\psi}^j_s\rbrace=0\ .
\end{gather}
The bosonic currents in the numerator can be split up
in $J^a$ for $a=1,\ldots,N^2-1$, $J^i$ and $\bar{J}^i$,
for $i=1,\ldots,N$, and $K$, where we define
\begin{equation}
J^i = t^A_{N+1,i}J^A\ ,
\qquad
\bar{J}^i = t^A_{i,N+1}J^A\ ,
\qquad
K = (N+1) \, t^A_{N+1,N+1} J^A\ .
\end{equation}
Here, $\bar{J}^i$ and $J^i$ correspond to the
$\mathbf{N}$ and $\mathbf{\bar{N}}$ of 
$\mathfrak{su}(N)$, respectively, while $K$
is the $\mathfrak{u}(1)$ current embedded into
$\mathfrak{su}(N+1)$. 
The $\mathfrak{u}(1)$ embedding into
$\mathfrak{so}(2N)$ can be written as 
\begin{equation}
j=-(N+1)(\psi^i\bar\psi^i)\ .
\end{equation}
The total $\mathfrak{u}(1)$ current is then
equal to $K+j$.
It will be useful to express the decoupled 
$\mathfrak{su}(N)_{k+1}$ currents in terms of the
decoupled $\mathfrak{su}(N+1)_{k}$ currents:
\begin{equation}
\widetilde{J}^a = J^a + t^a_{ij}(\psi^i \bar{\psi}^j)\ ,
\end{equation}
where we have assumed, without loss of generality, that 
the matrices $t^A$ for $A=N^2,\ldots,N(N+2)-1$
are of the form
\begin{equation}
t^A =\left(
\begin{array}{ccc|c}
&&&\\
\hphantom{0_N}&0_N&\hphantom{0_N}&\,\ast\\
&&&\\
\hline
&\ast&&\,0
\end{array}\right)\ , \qquad A=N^2,\ldots,N(N+2)-1\ ,
\end{equation}
and that $t^{N(N+2)}$ is diagonal.
We also define the unique spin-1 primary of the coset,
which is also the lowest field in the superconformal algebra, as
\begin{equation}
J = \frac{1}{N+k+1}\left(K-\frac{k}{N+1}\,j\right)\ .
\end{equation}
Then the stress-energy tensor of the coset theory
is given by the difference of the numerator
and denominator Sugawara tensors:
\begin{align}
T &= T_{\mathfrak{su}(N+1)}-T_{\mathfrak{su}(N)}
-T_{\mathfrak{u}(1)}+T_\text{free fermions}\nonumber\\
&=\frac{1}{2(N+k+1)}\Biggl((J^i\bar{J}^i)+(\bar{J}^iJ^i)
+k\left((\partial\psi^i\bar\psi^i)-
(\psi^i\partial\bar\psi^i)\right)\nonumber\\
 &\hspace{3.5cm} {}-2\,t^a_{ij}\left(J^a(\psi^i\bar\psi^j)\right)
-\frac{2}{N(N+1)}\,(Kj)\Biggr)\ ,
\end{align}
where we have used that
\begin{equation}
(J^AJ^A) = (J^aJ^a)+(J^i\bar{J}^i)+(\bar{J}^iJ^i)
+\frac{1}{N(N+1)}\,(KK)\ .
\end{equation}
We can split up the stress-energy tensor into three
mutually commuting stress-energy tensors given by
\begin{align}
T_{\rm b} &= \frac{1}{2(N+k+1)}\Biggl(
(J^i\bar{J}^i)+(\bar{J}^iJ^i)-\frac{1}{N+k}(J^aJ^a)
-\frac{1}{Nk}(KK)\Biggr)\ ,\nonumber\\
T_{\rm f} &= \frac{k}{2(N+k+1)}\Biggl(
(\partial\psi^i\bar\psi^i)-
(\psi^i\partial\bar\psi^i)
-\frac{2}{k}t^a_{ij}\left(J^a(\psi^i\bar\psi^j)\right)
+\frac{1}{k(N+k)}(J^aJ^a)\nonumber\\
&\qquad\qquad\qquad\qquad-\frac{1}{N(N+1)^2}
(jj)\Biggr)\ ,\nonumber\\
T_{{\rm (JJ)}} &= \frac{N+k+1}{2Nk}(JJ) \ ,
\end{align}
with central charges
\begin{align}
c_{\rm b} &=\frac{N(k-1)(N+2k+1)}{(N+k)(N+k+1)}\ ,\nonumber\\
c_{\rm f} &=\frac{k(N-1)(k+2N+1)}{(N+k)(N+k+1)}\ ,\nonumber\\
c_{\rm (JJ)}&=1\ ,
\end{align}
such that the total stress energy tensor reads
\begin{equation}\label{eq:T}
T=T_{\rm b}+T_{\rm f}+T_{\rm (JJ)}
\end{equation}
with total central charge
\begin{equation}
c =c_{\rm b}+c_{\rm f}+c_{\rm (JJ)}=\frac{3Nk}{N+k+1}\ .
\end{equation}
There is another elementary primary field
of conformal dimension $2$, which was called
$W^{20}$ in \cite{Candu:2012tr}.
We can make an ansatz
\begin{equation}\label{eq:w20}
W^{20} = \alpha \,T_{\rm b} + \beta\, T_{\rm f}
+\gamma\, T_{\rm (JJ)}\ .
\end{equation}
From the analysis in \cite{Candu:2012tr},
we know that $W^{20}$ satisfies the OPE
\begin{equation}
W^{20} \star W^{20} \sim n_2\, 1 + c_{22,2}\, W^{20}
+\frac{4\,n_2}{c-1}\left(T-\frac{3}{2c}\,(JJ)\right)\ .
\end{equation}
Demanding this as well as a vanishing central term
in the OPE $T \star W^{20}$, we obtain
\begin{align}\label{eq:w20_coeff}
\alpha&=-\sqrt{\frac{2k (N-1)(2N+k+1)(N+k+1)\,n_2 }{N(k-1) (N+2 k+1)(3N k-(N+k+1))}}\ ,\nonumber\\
\beta &=-\frac{N(k-1)(N+2k+1)}{k(N-1)(k+2N+1)}\,\alpha\nonumber\\
&=\sqrt{\frac{2N(k-1)(N+2k+1)(N+k+1)\,n_2}{k(N-1)(2N+k+1)(3 N k-(N+k+1))}}\ ,\nonumber\\
\gamma &= 0\ .
\end{align}
This then also reproduces correctly the form of 
$(c_{22,2})^2$ as predicted by eq.~(3.27) of \cite{Candu:2012tr}.
For the normalisation of $W^{20}$ we choose the convention
\begin{equation}
n_2 = -\frac{c}{6}(\nu+3)(\nu-3)\ ,
\end{equation}
where
\begin{align}
\nu &= 2\mu-1 = \frac{N-k-1}{N+k+1}\ ,\nonumber\\
c &= \frac{3Nk}{N+k+1}\ .
\end{align}
In the $c\rightarrow\infty$ limit, the parameters then become 
\begin{equation}\label{eq:alphabeta}
\alpha \to -\frac{\nu+3}{2\sqrt{3}}\ ,\qquad
\beta \to -\frac{\nu-3}{2\sqrt{3}}\ .
\end{equation}

\pagebreak

\end{document}